\input harvmac 
\input epsf.tex

\overfullrule=0mm

\newcount\figno
\figno=0
\def\fig#1#2#3{
\par\begingroup\parindent=0pt\leftskip=1cm\rightskip=1cm\parindent=0pt
\baselineskip=11pt
\global\advance\figno by 1
\midinsert
\epsfxsize=#3
\centerline{\epsfbox{#2}}
\vskip 12pt
{\bf Fig. \the\figno:} #1\par
\endinsert\endgroup\par
}
\def\figlabel#1{\xdef#1{\the\figno}}
\def\encadremath#1{\vbox{\hrule\hbox{\vrule\kern8pt\vbox{\kern8pt
\hbox{$\displaystyle #1$}\kern8pt}
\kern8pt\vrule}\hrule}}


\def\IR{\relax{\rm I\kern-.18em R}}
\font\cmss=cmss10 \font\cmsss=cmss10 at 7pt

\font\cmss=cmss10 \font\cmsss=cmss10 at 7pt
\def\IZ{\relax\ifmmode\mathchoice
{\hbox{\cmss Z\kern-.4em Z}}{\hbox{\cmss Z\kern-.4em Z}}
{\lower.9pt\hbox{\cmsss Z\kern-.4em Z}}
{\lower1.2pt\hbox{\cmsss Z\kern-.4em Z}}\else{\cmss Z\kern-.4em Z}\fi}
\def\IN{\relax{\rm I\kern-.18em N}}


\Title{\vbox{\hsize=3.truecm \hbox{SPhT/01-042}}}
{{\vbox {
\bigskip
\centerline{Critical and Multicritical Semi-Random}
\centerline{$(1+d)$-Dimensional Lattices and}
\centerline{Hard Objects in $d$ Dimensions}
}}}
\bigskip
\centerline{P. Di Francesco\foot{philippe@spht.saclay.cea.fr} and
E. Guitter\foot{guitter@spht.saclay.cea.fr}}
\medskip
\centerline{ \it CEA-Saclay, Service de Physique Th\'eorique,}
\centerline{ \it F-91191 Gif sur Yvette Cedex, France}
\medskip

\vskip .5in

\noindent

We investigate models of $(1+d)$-D Lorentzian semi-random lattices with
one random (space-like) direction and $d$ regular (time-like) ones.
We prove a general inversion formula expressing the partition
function of these models as the inverse of that of hard
objects in $d$ dimensions. This allows for an exact solution
of a variety of new models including critical and multicritical
generalized $(1+1)$-D Lorentzian surfaces, with fractal 
dimensions $d_F=k+1$, $k=1,2,3,...$, as well as a new
model of $(1+2)$-D critical 
tetrahedral complexes, with fractal dimension $d_F=12/5$. 
Critical exponents and 
universal scaling functions follow from this solution.  
We finally establish a general connection between 
$(1+d)$-D Lorentzian lattices and
directed-site lattice animals in $(1+d)$ dimensions. 

\noindent
\Date{04/01}

\nref\DGZ{P. Di Francesco, P. Ginsparg
and J. Zinn--Justin, {\it 2D Gravity and Random Matrices},
Physics Reports {\bf 254} (1995) 1-131.}
\nref\ADJ{J.\ Ambj\o rn, B.\ Durhuus and T.\ Johnsson, {\it Quantum
Geometry}, Cambridge University Press, 1997.}
\nref\AL{J. Ambj\o rn and R. Loll, {\it Non-perturbative Lorentzian 
Quantum Gravity, Causality and Topology Change}, 
Nucl. Phys. {\bf B536 [FS]} (1998) 407, hep-th/9805108.}
\nref\FGK{P. Di Francesco, E. Guitter and C. Kristjansen, 
{\it Integrable 2D Lorentzian Gravity and Random Walks}, Nucl. Phys. 
{\bf B567 [FS]} (2000) 515, hep-th/9907084.}
\nref\AMBJ{J. Ambj\o rn,  J. Jurkiewicz and R. Loll,
{\it Non-perturbative 3d Lorentzian Quantum Gravity}, preprint
hep-th/0011276.}
\nref\FGKBIS{P.\ Di Francesco, E.\ Guitter and C.\ Kristjansen, {\it
Generalized Lorentzian Gravity in 1+1D and the Calogero Hamiltonian},
preprint hep-th/0010259 (2000).}
\nref\BAXHH{R. J. Baxter, {\it Hard hexagons: exact solution}, J. Phys. {\bf A 13}
(1980) L61-L70; R. J. Baxter and S.K. Tsang, {\it Entropy of Hard Hexagons},
J. Phys. {\bf A 13} (1980) 1023-1030; see also
R. J. Baxter, {\it Exactly solved models in statistical mechanics},
Academic Press, London (1984).}
\nref\CARD{J. Cardy, {\it Directed lattice animals and the 
Lee-Yang edge singularity},  J. Phys. {\bf A 15} (1982) L593-L595.}
\nref\DHAR{D. Dhar, {\it Equivalence of two-dimensional
directed-site animal problem to Baxter's hard-square lattice gas model},
Phys. Rev. Lett. {\bf 49} (1982) 959-962;
{\it Exact solution of a directed-site animals-enumeration problem
in three dimensions}, Phys. Rev. Lett. {\bf 51} (1982) 853-856.} 
\nref\BIJO{D. Bisch and V. Jones, {\it Algebras associated to intermediate
subfactors}, Inv. Math. {\bf 128} (1997) 89.}
\nref\PATA{W. Lang, {\it On generalizations of the Stirling number triangles}, 
J. Integer Seqs., {\bf Vol. 3} (2000), $\#$00.2.4 and references therein.}
\nref\BAT{H. Bateman, {\it Higher Transcendental Functions}, Vol. III,
McGraw-Hill (1953).}
\nref\KF{D. Kurze and M. Fisher, {\it Yang-Lee edge singularities
at high temperatures}, Phys. Rev. {\bf B20} (1979) 2785.}
\nref\DSZ{P. Di Francesco, H. Saleur and J.-B. Zuber, 
{\it Generalized Coulomb gas formalism for two-dimensional critical
models based on SU(2) coset construction}, Nucl. Phys. {\bf B300
[FS]} (1988) 393-432.}
\nref\CARTWO{J. Cardy, {\it Conformal invariance and the Yang-Lee edge
singularity in two dimensions}, 
Phys. Rev. Lett. {\bf 54}, No. 13 (1985) 1354-1356.}
\nref\BAX{R. J. Baxter, {\it Planar lattice gases with nearest-neighbour
exclusion}, preprint cond-mat/9811264 (1998).}
\nref\ABF{G. Andrews, R. Baxter and P. Forrester, 
{\it Eight vertex SOS model and generalized Rogers-Ramanujan type identities},
J. Stat. Phys.  {\bf 35} (1984) 193-266.}
\nref\TROISD{M. Bowick, P. Di Francesco,
O. Golinelli and E. Guitter, {\it Three-dimensional Folding of the Triangular
Lattice}, Nucl. Phys. {\bf B450[FS]} (1995) 463-494, cond-mat/9502063.}
\nref\VIE{G. Viennot, {\it Heaps of pieces I:
basic definitions and combinatorial lemmas},
in G. Labelle and P. Leroux, Eds, {\it Combinatoire
\'enum\'erative}, Lect. Notes in Math. {\bf 1234} (1986)
321-346.}
\nref\BP{J. B\'etr\'ema and J. Penaud, { Mod\`eles avec
particules dures, animaux dirig\'es, et s\'eries
en variables partiellement commutatives}, Labri Report (1994),
http://dept-info.labri.u-bordeaux.fr/~betrema.}
\nref\BM{M. Bousquet-M\'elou and A. Rechnitzer, {\it Lattice animals and
heaps of dimers}, Labri preprint (2000),
http://dept-info.labri.u-bordeaux.fr/~bousquet/publis.html.}


\newsec{Introduction}

The study of random lattices is an important subject with relations to many
areas of physics such as quantum gravity or fluid membranes. 
One of the most efficient descriptions of such random lattices
is through matrix models which generate discrete two-dimensional
random surfaces in the
form of tesselations made of tiles of arbitrary valences
(for reviews see \DGZ\-\ADJ\ and references therein). The archetypical example 
is random triangulations perused in the context of 2D quantum gravity. 
Analogous tesselations can be considered in higher dimension but unfortunately
no such powerful tool as matrix models is available, making the subject quite
difficult.
Recently a new type of random lattices was introduced \AL\ referred to as
Lorentzian random lattices in which one particular (time) direction is regular
while the other (space) ones are random. This allows to view a, say, 
$(D+1)$-dimensional
Lorentzian lattice as the time evolution of a $D$-dimensional 
usual random lattice.
More generally, we may as well consider semi-random
lattices, with a number, say, 
$D$ of random  directions and $d$ of regular ones. 
We shall refer to these lattices as $(D+d)$-D Lorentzian lattices.
These can be viewed as interpolations between regular lattices ($D=0$)
and fully random ones ($d=0$).
By a slight abuse of language, we shall refer to the $d$ regular
directions as ``time directions" and to the $D$ random ones 
as ``space directions", 
although our analysis is purely statistical (no dynamics).
The cases considered so far correspond to $d=1$ and
$D=1$ \AL\-\FGK\ or $2$ \AMBJ. 

In the present work, we focus on the case of $D=1$ and
arbitrary $d\geq 1$, referred to as $(1+d)$-D Lorentzian-type lattices in the following.
For such models, we will derive a powerful {\it inversion relation}, expressing the
partition function for such a $(1+d)$-D Lorentzian lattice as the inverse of 
the partition function 
of some hard object model on the time-like $d$-dimensional regular lattice.
More precisely, this relation takes the general form
\eqn\forgeninv{ Z_{1+d}(\{t_i\}) = {1\over Z_d^{h}(\{-t_i\}) }}
where $Z_{1+d}(\{t_i\})$ is the partition function of our $(1+d)$-D Lorentzian lattices
with activities $t_i$ per tile of type $i$, and $Z_d^{h}(\{-t_i\})$
denotes the partition function of hard objects on the time-like $d$-dimensional
regular lattice with activities $-t_i$ per hard object of corresponding type $i$.
This relation \forgeninv\ is very general, and holds for any dimension $d$ and
any {\it fixed} $d$-dimensional (but not necessarily regular) time lattice,
any type of tile $i$, and even with time-dependent activities $t_i$. 
We may view the inversion relation \forgeninv\ as some type of ``boson-fermion"
correspondence, relating the partition function of weakly interacting bosons,
namely the tiles of the lattice, to the inverse of that of locally interacting
fermions, namely the hard objects with nearest neighbor exclusion. 

As a first application, we show how to relate the partition function of 
$(1+1)$-D Lorentzian
triangulations to that of the hard dimer model on a line, which allows for
a very simple solution of the problem in terms of a $2\times 2$ transfer matrix,
as opposed to the previous solutions relying on transfer matrices of infinite size
\AL\ \FGK\ \FGKBIS. This new approach allows to build and solve many more 
models of $(1+1)$-D  
Lorentzian surfaces, such as those made of larger ($2(i+1)$-gonal) tiles,
in connection with hard multimers on a line. In particular, we are able to reach 
new multicritical points for these surfaces, 
displaying new large scale universal properties, and fractal dimension $D_F=k+1$,
$k=1,2,3,...$.

Going to higher dimension, we then
introduce a model of $(1+2)$-D Lorentzian tetrahedral complexes, i.e.
semi-random lattices made of tetrahedra, and apply our
inversion relation to express its 
partition function in terms of that of 2D hard hexagons solved by 
Baxter \BAXHH. As an outcome, we immediately
obtain the large scale behavior of these new semi-random lattices.   

Analogous relations between $(1+d)$-dimensional problems and $d$-dimensional
nearest-neighbor exclusion models have already been found in the context
of directed-site lattice animal enumeration (DSAE) problems \CARD\-\DHAR. 
This suggests
the existence of a connection between Lorentzian-type $(1+d)$-D Lorentzian 
lattices and $(1+d)$-dimensional DSAE. We will establish
such a connection in which animals will appear as a particular subclass of 
Lorentzian lattices.
Our inversion relation actually provides an alternative and more direct derivation
of the equivalence between DSAE and nearest neighbor exclusion models. 

The paper is organized as follows. In Section 2, we discuss the case of 
Lorentzian $(1+1)$-D surfaces. We first derive the inversion formula by focusing
on the simplest case of $(1+1)$-D Lorentzian triangulations (Section 2.1).
This allows us to rederive very simply some of the known properties
of these surfaces. In Section 2.2, we extend the inversion formula to include
$(1+1)$-D Lorentzian surfaces made of time-like $2(i+1)$-gons with activity
$t_i$ $i=1,2,3...$, now corresponding 
to hard $(i+1)$-mers on a line, and derive the corresponding thermodynamic
partition function and loop-loop propagator. 
The first application concerns the case of surfaces made of only one type
of such tiles (say $2(k+1)$-gons, with fixed $k$) and is discussed in 
Section 2.3. Next we show in Section 2.4 how to obtain multicritical models
by fine-tuning the various activities $t_i$. For these models, we 
compute the corresponding scaling exponents as well as universal scaling
functions. 
Section 3 is devoted to the study of $(1+2)$-D Lorentzian 
tetrahedral complexes. We first define the model in Section 3.1 in terms
of plaquettes living in tubes of hexagonal section. We then
apply in Section 3.2 the inversion relation to obtain the critical behavior
of our model in terms of that of hard hexagons at the Lee-Yang
edge singularity point.  
In Section 4, we make the connection between our models and 
directed-site lattice animals, both in the $(1+1)$-D case (Section 4.1) and
in the $(1+2)$-D one (Section 4.2). Finally,
Section 5 is devoted to a discussion of the (infinite)
transfer matrices for our models. 
We first derive in Section 5.1 the various $(1+1)$-D transfer matrices in 
terms of the corresponding finite ones for hard objects
on a line. The most technical cases of this discussion are treated 
in Appendix A. We then use in Section 5.2 the equivalence to hard objects
to construct more general parametric families of mutually commuting  
transfer matrices corresponding to integrable models containing our
semi-random lattice models as particular points.
We gather a few concluding remarks in Section 6.

\newsec{Critical and Multicritical Models of Lorentzian $(1+1)$-D Surfaces}

\subsec{Inversion principle: $(1+1)$-D Lorentzian triangulations vs 1D hard dimers}

In this section, we introduce the fundamental inversion formula relating 
the partition functions of Lorentzian-type semi-random lattices 
and hard objects in one less dimension. For simplicity, we specialize here to the 
simplest model of pure Lorentzian triangulations in $(1+1)$-D, which corresponds
to hard dimers on a line.

\fig{A typical $(1+1)$-D Lorentzian triangulation together with its dual,
made of $T$ time-slices. The triangulation is regular in the ``time"
direction and random in the ``space" direction with an arbitrary succession
of up and down triangles in each slice. Triangles of neighboring slices
may be paired so as to form time-like lozenges, 
such as the shaded one in the figure. These elementary building blocks 
translate into vertical edges in the dual picture.}{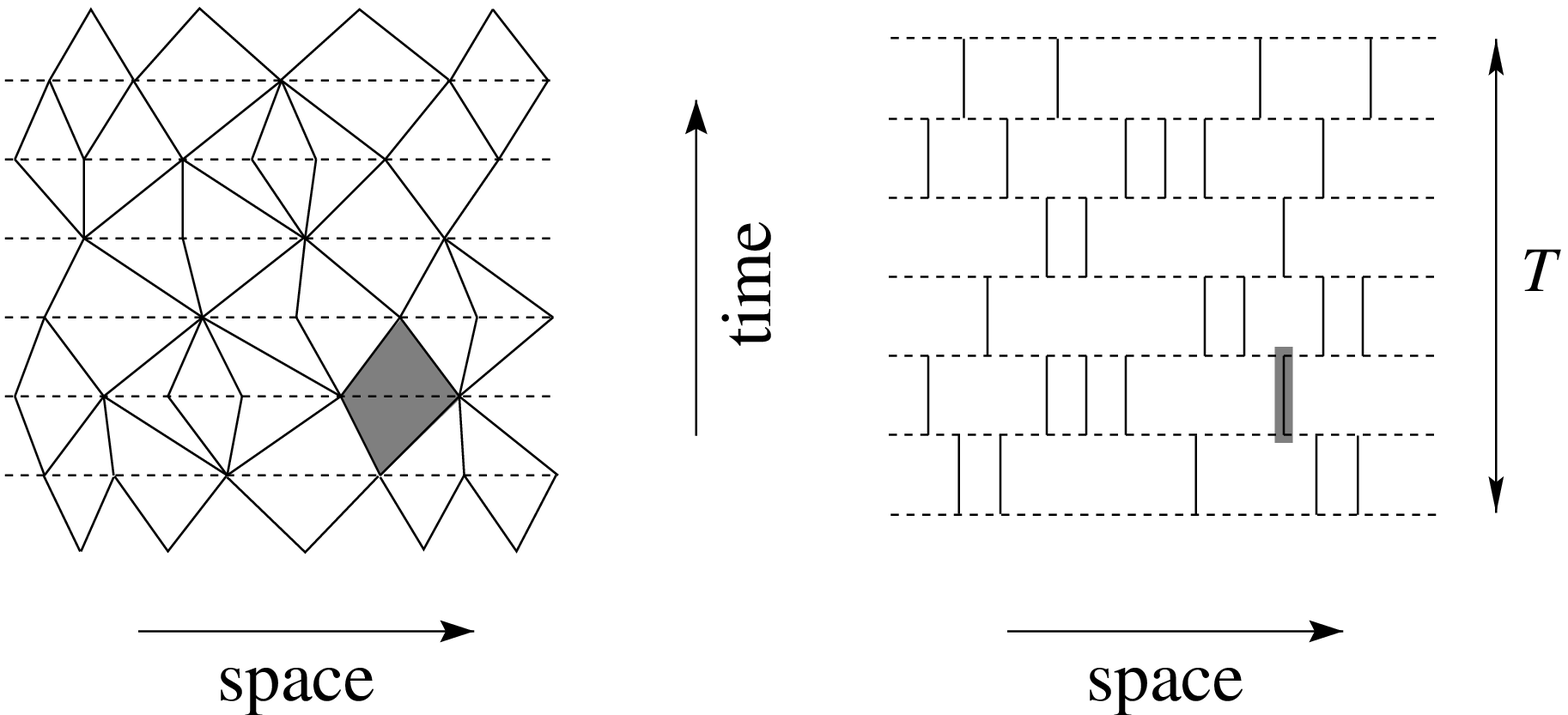}{14.truecm}
\figlabel\lorgra

We start with the partition function $Z_T(t)$ of Lorentzian triangulations 
\AL\ with
$T$ time slices and an activity $t$ per pair of neighboring triangles sharing
a space-like edge (see Fig.\lorgra). 
The corresponding ``time-like" lozenges form the  building
blocks in the construction of the surfaces. In the dual picture, they translate
into time-like (vertical) edges  connecting successive space-like 
(horizontal) lines.

\fig{A configuration of $(1+1)$-D Lorentzian triangulation in the dual picture, together
with its left vertical projection. The latter is obtained by letting the 
foreground of the triangulation, made of the leftmost
edges, slide horizontally all the way to the vertical line on the left.
This projection clearly defines a hard dimer configuration on the vertical 
line.}{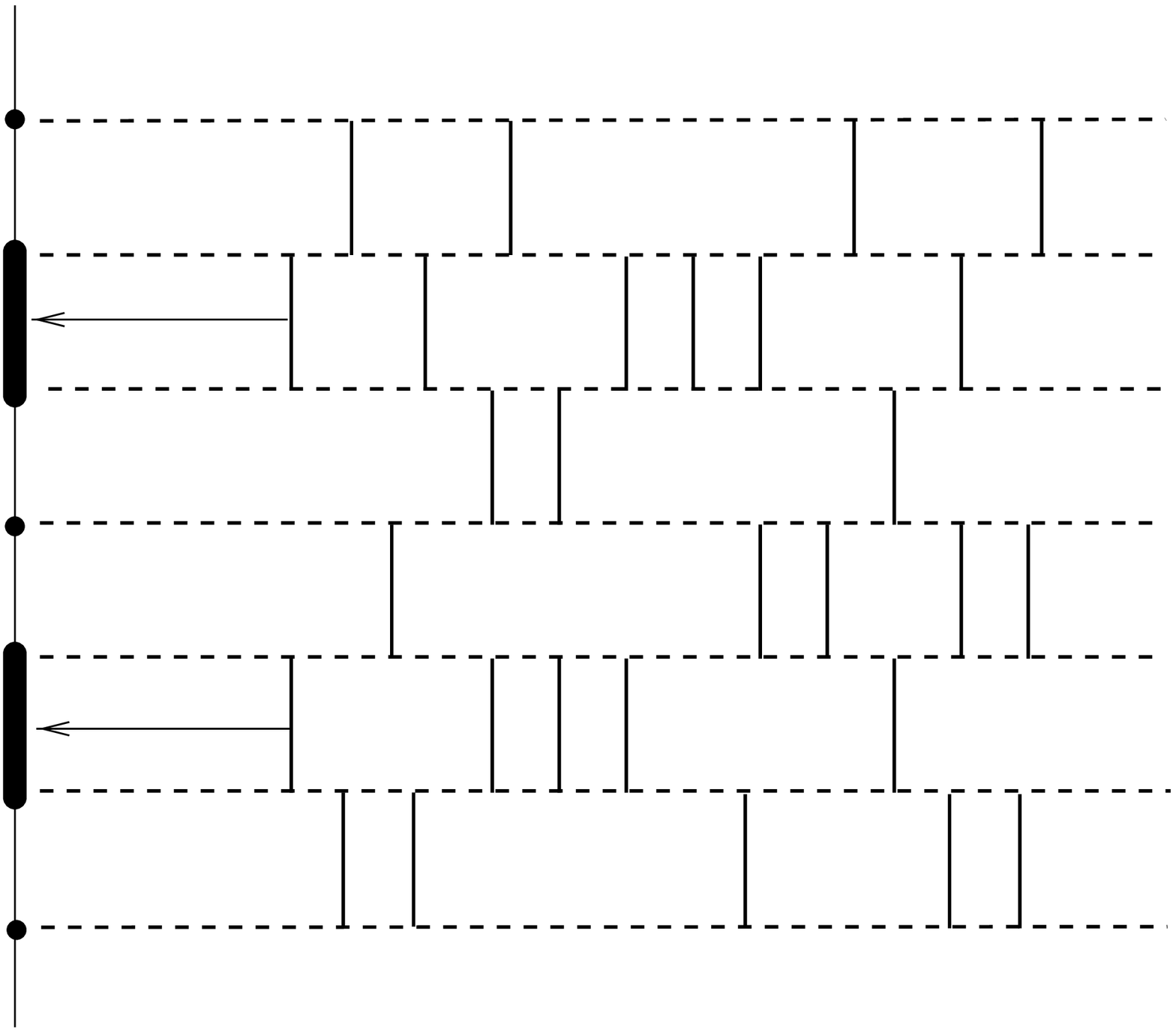}{8.truecm}
\figlabel\hd

The idea behind the correspondence to hard dimers is to decompose the 
surface configurations according to their left vertical projection defined
as follows. 
Let us allow for the vertical edges to slide along the horizontal direction
without passing one-another both within the same time-slice and between two
consecutive ones, thus preserving the relative positioning between them.
We then single out those vertical edges which can be taken all the way to 
a vertical line on the left  without moving the others. The resulting
configuration of edges along this vertical line constitutes
the left vertical projection of our surface. Clearly, it defines a
hard dimer configuration of the vertical integral segment $[0,T]\subset \IZ$ 
as depicted in Fig.\hd.
To get dimers we simply view each edge as linking its two endpoints.
The hardness simply means that any integer point in $[0,T]$ belongs to
{\it at most} one dimer, translating into a mutual avoidance of dimers.
This is due to the fact that on any two consecutive time-slices,
no more than one edge can be projected.
  
We therefore write the partition function $Z_T(t)$ as
\eqn\zeg{ Z_T(t)=\sum_{{\rm hard}\ {\rm dimer}\ {\rm config.}\ C}
t^{|C|} \, Z_T^{(C)}(t) }
where the sum extends over all hard dimer configurations $C$ on the
integer segment $[0,T]$ (including the empty one), and where $Z_T^{(C)}(t)$
is the restricted partition function involving those configurations having
projection $C$, from which we have factored out the weight $t^{|C|}$
of the projected part, $|C|$ denoting the number of dimers in $C$.
More generally we have the relations

\fig{By completing an arbitrary Lorentzian triangulation with a hard dimer
configuration $D$ (a), 
we build a larger triangulation (b) whose projection $C$ contains 
$D$ (c). With this procedure (with fixed $D$) we build exactly once
all Lorentzian triangulations whose projection $C$ contains $D$.}{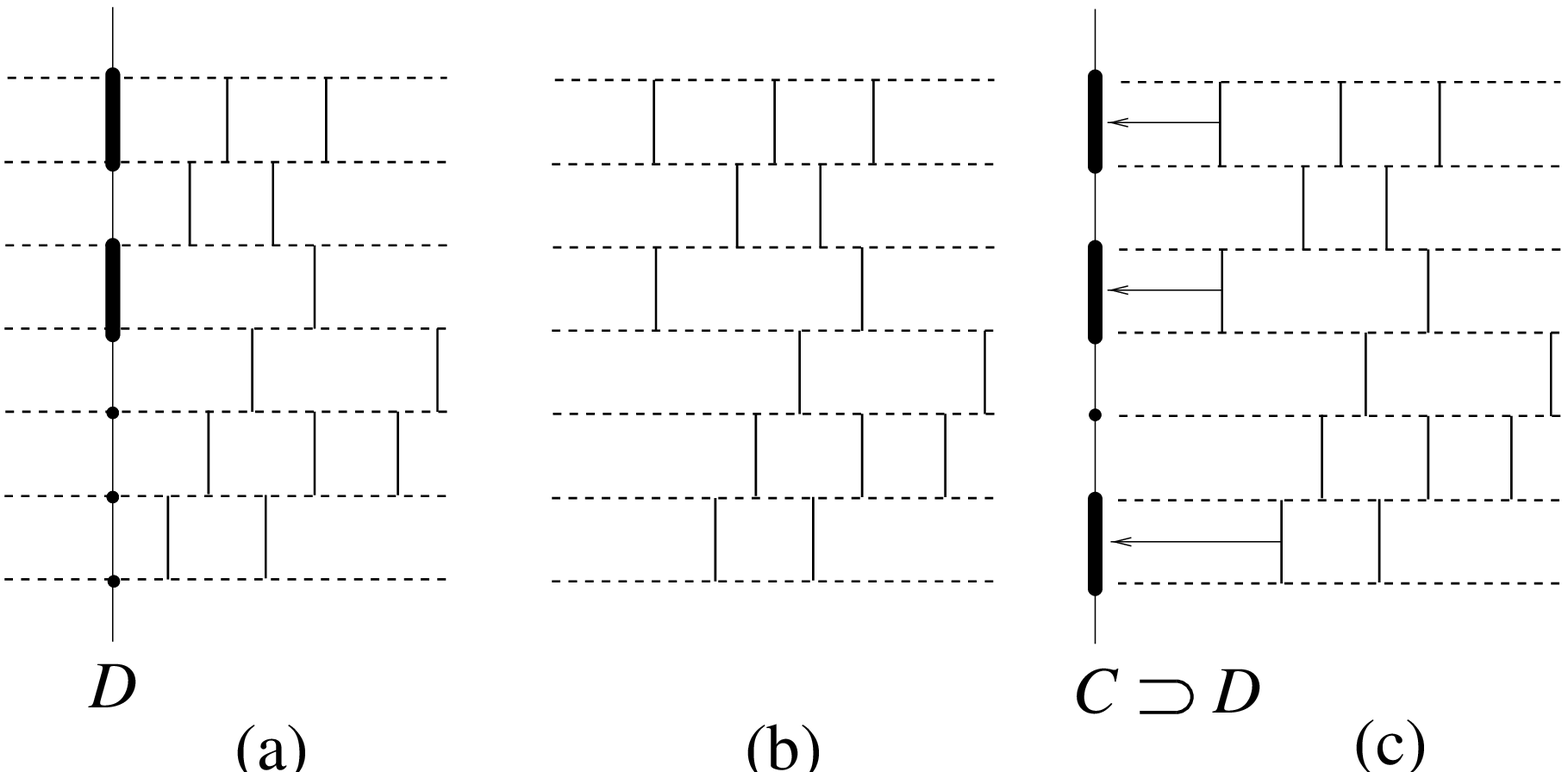}{14.truecm}
\figlabel\proj

\eqn\zegplus{ t^{|D|} Z_T(t)=\sum_{C \supset D} 
t^{|C|} \, Z_T^{(C)}(t) }
valid for any hard dimer configuration $D$ (eqn.\zeg\ corresponding to $D=\emptyset$).
This expresses the fact that by completing any configuration of Lorentzian
triangulation with $T$ time slices (in the dual picture) by a given 
left column of vertical edges (corresponding to a hard dimer
configuration $D$), one builds exactly once each configuration
having a projection containing $D$, i.e. having $D$ as a sub-configuration
(see Fig.\proj).  
This latter relation is easily inverted using the celebrated M\"obius 
inversion\foot{ Note that the M\"obius inversion involves usually the opposite
order relation $\subset$ but it works similarly for the order relation
$\supset$ used here.}  
formula, leading to
\eqn\invzeg{ t^{|C|} \, Z_T^{(C)}(t) = \sum_{D \supset C} (-1)^{|D|-|C|} 
t^{|D|} Z_T(t) }
Noting that $Z_T(t)$ factors out of the sum on the rhs, we finally get
\eqn\forZ{ Z_T(t)= {(-t)^{|C|} \, Z_T^{(C)}(t) \over \sum\limits_{D \supset C} (-t)^{|D|}}}
Picking $C=\emptyset$, we arrive at our fundamental inversion relation
\eqn\funda{ Z_T(t)= { 1\over Z_T^{hd}(-t)} }
where 
\eqn\hardi{Z_T^{hd}(z)= \sum_{{\rm hard}\ {\rm dimer}\ {\rm config.}\ D}
z^{|D|} }
denotes the standard partition function for hard dimers with fugacity $z$ per dimer. 
As we already mentioned in the introduction,
this relation is a generalization of the boson-fermion correspondence relating 
for instance the
partition function $1/(1-t)$ of free bosons with fugacity $t$ per particle on a point
to the inverse of $1+z=1-t$ of that of a fermion with fugacity $z=-t$ on a point.

The formula \forZ\ also implies, upon substituting \hardi\ that
\eqn\corr{ Z_T^{(C)}(t) = { Z_{[0,T]\setminus C}^{hd}(-t) \over Z_T^{hd}(-t)}}
where
\eqn\hadi{ Z_{[0,T]\setminus C}^{hd}(z) = \sum_{D \supset C} (z)^{|D|-|C|}}
is the hard dimer partition function on the segment $[0,T]$ minus the occupied
points of $C$.  

The above construction is very general. In particular, our inversion formula
\funda\ holds also if we attach an extra fugacity $w_s$ per edge
inside the 
time-slice $s$ resulting in a total weight
$z_s=-t w_s$ per dimer in the slice $s$ in the hard dimer language, 
allowing in particular to recover the loop-loop propagator
(partition function with a fixed number of triangles at times $0$ and $T$),
and correlation functions for the numbers of triangles in given slices. 
It will also be extended in Section 2.2 below to the case of 
surfaces made of larger building blocks corresponding to hard multimers on
a line. Finally in Section 3 we will extend it to higher-dimensional
semi-random lattices. 

With this construction we have reduced the $(1+1)$-D Lorentzian gravity-type
problems to that, much simpler, of hard objects on a line.

To complete this section, let us re-derive the partition function 
of pure Lorentzian triangulations $Z_T(t)$ from the hard dimer
equivalence. The partition function $Z_T^{hd}(z)$ is easily
computed by use of a $2\times 2$ transfer
matrix $\cal T$ 
between successive
segments in either empty or occupied states:
\eqn\tmathd{Z_T^{hd}(z)=v^t{\cal T}^{T+1}v=\pmatrix{1 & 0\cr} 
\pmatrix{1 & 1 \cr z & 0 \cr}^{T+1}\pmatrix{1\cr 0\cr} }
We may read directly from this the asymptotic large $T$ behavior of $Z_T(t)$. 
Indeed, \tmathd\ is dominated for large $T$ by the largest eigenvalue 
$\lambda_+$ of $\cal T$, satisfying
of $\lambda^2-\lambda-z=0$ (with solutions 
$\lambda_\pm(z)=(1\pm\sqrt{1+4z})/2$). Therefore
we find that for large $T$ 
\eqn\tmathd{Z_T(t)={1\over Z_T^{hd}(-t)}\sim \mu^T(t)\ \ 
{\rm where}\ \ \mu(t)={1\over \lambda_+(-t)}}
is the {\it smallest} solution of 
\eqn\charaone{ \mu=1+t\mu^2 }
namely the celebrated generating function of the Catalan numbers
$\mu(t)=(1-\sqrt{1-4t})/(2 t)$.
More precisely, we have
\eqn\hdonepart{Z_T^{hd}(z)= {\lambda_+(z)^{T+2} -\lambda_-(z)^{T+2} \over
\lambda_+(z)-\lambda_-(z)}}
leading to 
\eqn\invhdpar{ Z_T(t)= {1-q^2 \over 1+q^2} {(1+q^2)^{T+2}\over 1-q^{2(T+2)}} }
where we have set $q=\sqrt{t} \mu(t)$ (or equivalently $1/\sqrt{t}=q+1/q$).
Expanding the denominator in \invhdpar\ we immediately read the eigenvalues
of the infinite transfer matrix for Lorentzian triangulations 
$(q+{1\over q})q^{2n+1}$,
$n=0,1,2,...$, thus recovering the harmonic oscillator 
eigenvalues \FGK\-\FGKBIS.

\fig{A typical Lorentzian triangulation with projection $C_0$,
made of a single dimer in the lowest position. This induces a
staircase-type boundary condition on the left side of the
triangulation, namely that any edge in a given time-slice must have at least one
edge on its left in the slice just below.}{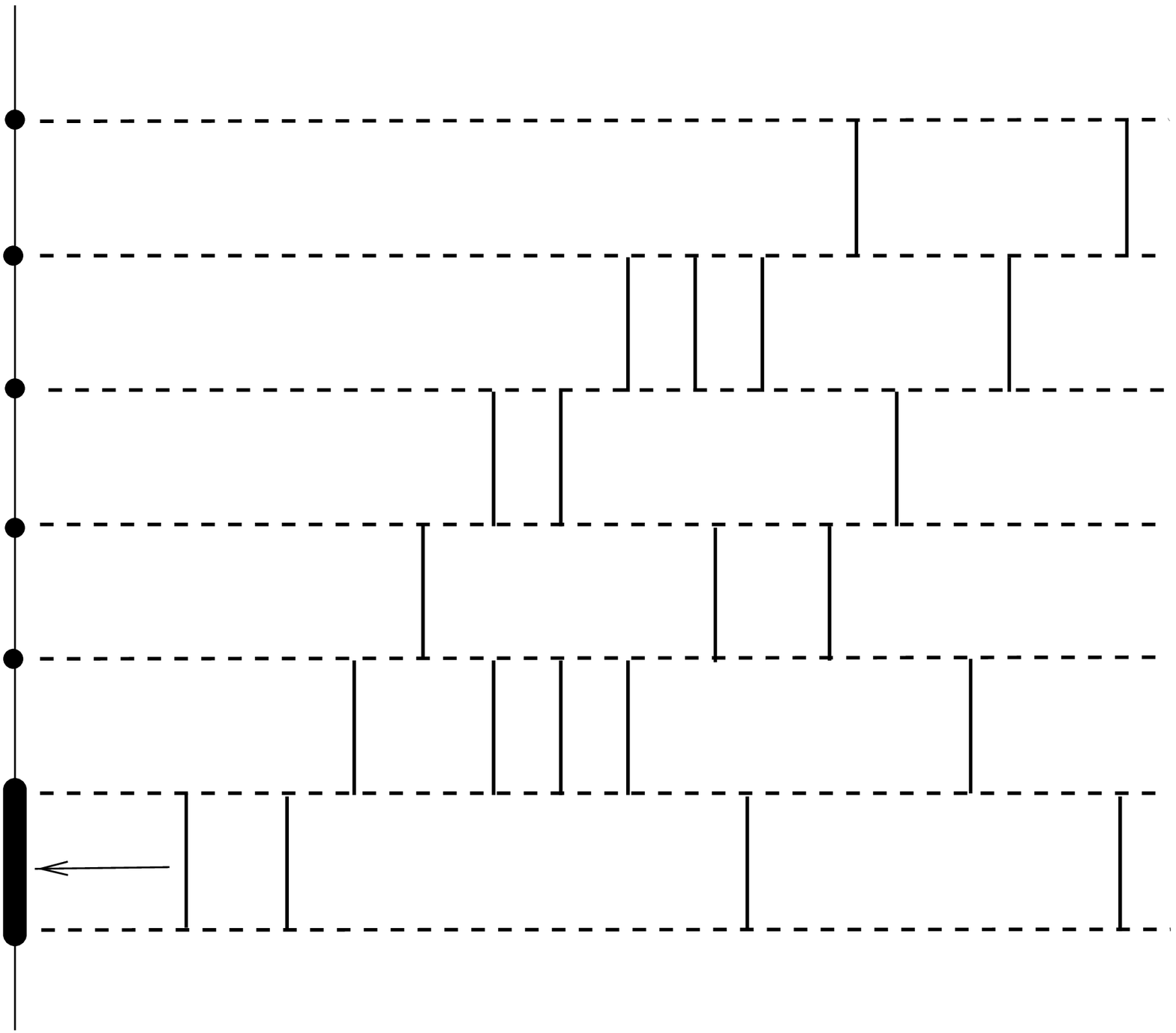}{6.5truecm}
\figlabel\lowerleft

As a final remark, we may interpret $\mu(t)$ as the partition function of 
a sub-class of triangulations on the semi-infinite time interval 
$[0,\infty)$. To get a finite partition function it is sufficient to
demand that the projection of the configurations is either the vacuum
or the hard dimer configuration $C_0$ made of a single dimer in the first 
slice (at time 0, see Fig.\lowerleft). 
This implies a so-called ``staircase" boundary condition on the
left of the configurations, namely that an edge occurs in the slice $s$ only if 
an edge already occured on its left in the slice $s-1$ below. Indeed, we
write this partition function as
\eqn\parft{\eqalign{ 1+ t \lim_{T\to \infty} Z_T^{(C_0)}(t) &= 1+t 
\lim_{T\to \infty}
{Z_{[0,T]\setminus C_0}^{hd}(-t) \over Z_{[0,T]}^{hd}(-t)}\cr 
&=1 +t \lim_{T\to \infty} {\mu^T \over \mu^{T-2}} =1+t \mu^2 = \mu \cr}}

\fig{The one-to-one correspondence between ``left staircase boundary" triangulations
and random walks from the origin to itself on the integer half-line.
We first rewrite the configuration of edges forming the triangulation as
a tree, by connecting each edge to that sitting just below on its left
(connections are represented by thick horizontal lines here).
By following the contour of the tree from the lower left branch to the lower
right, the sequence of ascents and descents gives rise to
the directed walk represented in thin solid line (to get a nice directed walk, we must
first place each vertical edge at a horizontal position equal to total number
of ascents and descents along the tree preceding it). Conversely,
each such walk gives rise to a unique tree, therefore to a unique triangulation
with the staircase boundary condition. This correspondence
includes the empty triangulation
with projection $\emptyset$, corresponding to 
the walk of $0$ step.}{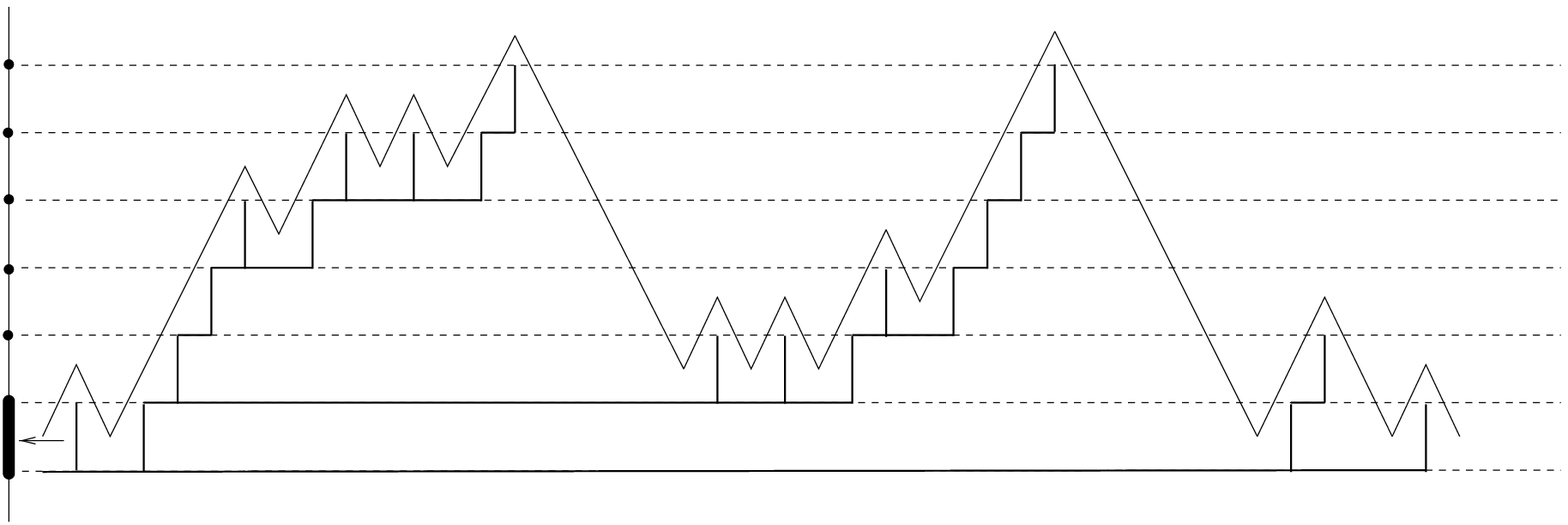}{16.truecm}
\figlabel\rw

As explained in Ref.\FGK\ and depicted in Fig.\rw,
the configurations above are in one-to-one correspondence with discrete 
Random Walks with steps of $\pm 1$ on the integral half-line $[0,\infty)$, 
starting and ending at
the origin, and with a weight $t$ per ascending step. 

As we will see in the following sections, all the above equivalences will nicely
generalize to more involved cases.

\subsec{Generalized $(1+1)$-D Lorentzian surfaces vs 1D hard multimers} 

\fig{A typical generalized $(1+1)$-D discrete Lorentzian surface (a) made
of time-like lozenges, hexagons, octagons, etc... respectively
weighted by $t_1,t_2,t_3,...$ Its dual (b) is made
of vertical edges of length $1,2,3,...$ extending over several time-slices.
The corresponding left vertical projection is nothing but a configuration 
of the hard multimer model on the integer segment $[0,T]$.}{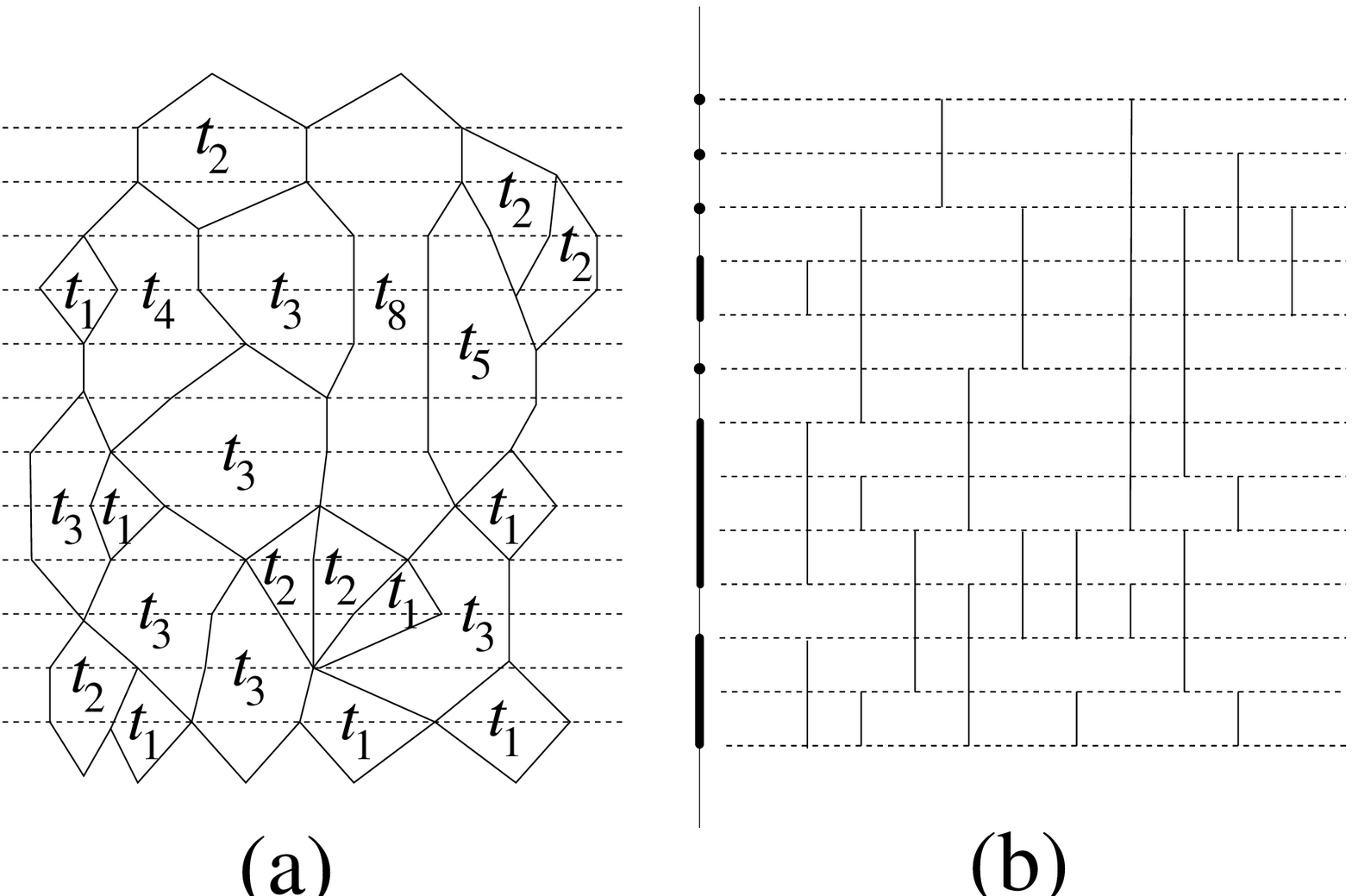}{14.truecm}
\figlabel\multimer

In this section, we introduce generalized discrete Lorentzian surfaces made
out of various tiles, including lozenges, hexagons, octagons, etc...
To be more precise, we wish to compute the partition function $Z_T(\{t_i\})$
of surfaces which in the dual picture look like
Fig.\multimer, with vertical edges of arbitrary length with a fugacity $t_i$
per edge of total length $i$, $i=1,2,3...$ and taken on a time segment $[0,T]$.
These edges are dual to particular time-like $2(i+1)$-gons (see Fig.\multimer\ (a)).
The inversion formula \funda\ generalizes straightforwardly to this case.
Firstly we now have the generalization of eqn. \zegplus\
\eqn\zegencoreplus{ \prod_i t_i^{n_i(D)} Z_T(\{t_i\})=\sum_{C \supset D}
\prod_i t_i^{n_i(C)}\, Z_T^{(C)}(\{t_i\}) }  
where $D$ (resp. $C$) denote hard multimer configurations on the integral segment
$[0,T]$ with $n_i(D)$ (resp. $n_i(C)$) ($i+1$)-mers. Again in such a configuration
hard multimers avoid one-another in that a given point may belong to at most one
multimer. By inclusion of configurations $C \supset D$ we mean that $C$ contains all
the multimers of $D$ plus possibly others. Eqns. \zegencoreplus\ are easily
inverted by the M\"obius inversion formula to finally yield:
\eqn\fundafin{ Z_T(\{t_i\})= {1\over Z_T^{hm}(\{z_i=-t_i\})} }
where $Z_T^{hm}(\{z_i\})$ denotes the partition function of hard multimers
on the integral segment $[0,T]$ with a fugacity $z_i$ per ($i+1$)-mer.

Again, in the hard multimer language we simply have to diagonalize
the corresponding transfer matrix. Let us fix for definiteness a maximal length $k$
for the edges, corresponding to a model of di-, tri-, ..., ($k+1$)-mers.
This truncates the transfer matrix to a size $(k+1)\times (k+1)$, 
where the $(k+1)$ possible states correspond to the empty state, the 
state occupied by the lowest monomer of a ($i\ge 1$)-mer, the state occupied
by the second lowest monomer of a ($i\ge 2$)-mer, ...
The transfer matrix then reads 
\eqn\transinf{ {\cal T}_k= \pmatrix{ 
1 & 1 & 1 & \cdots & 1 & 1\cr 
z_1 & 0 & 0 & \cdots & 0 & 0\cr
0 & {z_2 \over z_1} & 0 & \cdots & 0 & 0 \cr
0 & 0 & {z_3 \over z_2} & \cdots & 0 & 0\cr
\vdots & \vdots & \ddots & \ddots & \ddots & \vdots\cr
0 & 0 & \cdots & \cdots & {z_k \over z_{k-1}} & 0\cr} }
and the partition function is expressed as
\eqn\pfmulti{Z_T^{hm}(\{z_i\})= v^t \big({\cal T}_k\big)^{T+1} v}
where $v^t=(1,0,0,...,0)$.
Again the large $T$ asymptotics of \fundafin\ are governed by the largest
eigenvalue $\lambda(\{z_i\})$ of ${\cal T}_k$, leading to
\eqn\mudef{ Z_T(\{t_i\})\sim \mu^T(\{t_i\}) }
where $\mu(\{t_i\})=1/\lambda(\{z_i=-t_i\})$ is the smallest (in module) solution of
the characteristic equation
\eqn\charmut{ \mu= 1+ \sum_{i=1}^k t_i \mu^{i+1} }
Here again, $\mu$ may be interpreted as the partition function for the particular
Lorentzian surfaces with semi-infinite time interval $[0,\infty)$, having  
a projection either empty (contribution of $1$ in the rhs of
\charmut), or equal to a single multimer of either length $i$,
extending over the $i$ first time slices, $i=1,2,...,k$ (contributions
of $t_i \mu^{i+1}$ in the rhs of \charmut).

\fig{The one-to-one correspondence between ``left staircase boundary" generalized
discrete Lorentzian surfaces 
and random walks from the origin to itself on the integer half-line, with ascending steps
of arbitrary length and descending steps of $-1$. As in the triangulation case,
we associate to each ``left staircase boundary" surface configuration
a tree connecting each edges to that sitting just below it on its left. 
The walk is then defined
as a walk along the tree, with the convention that it makes an ascending
step of $+i$ when going up along an $(i+1)$-mer, whereas it only makes
descending steps of $-1$ (again to best represent it as a directed walk,
we must place the edges at horizontal positions equal to the total number
of ascents and descents preceding them). 
Conversely any such walk gives rise to a unique tree
with branches of arbitrary integer lengths
and therefore a unique ``left staircase boundary" surface.}{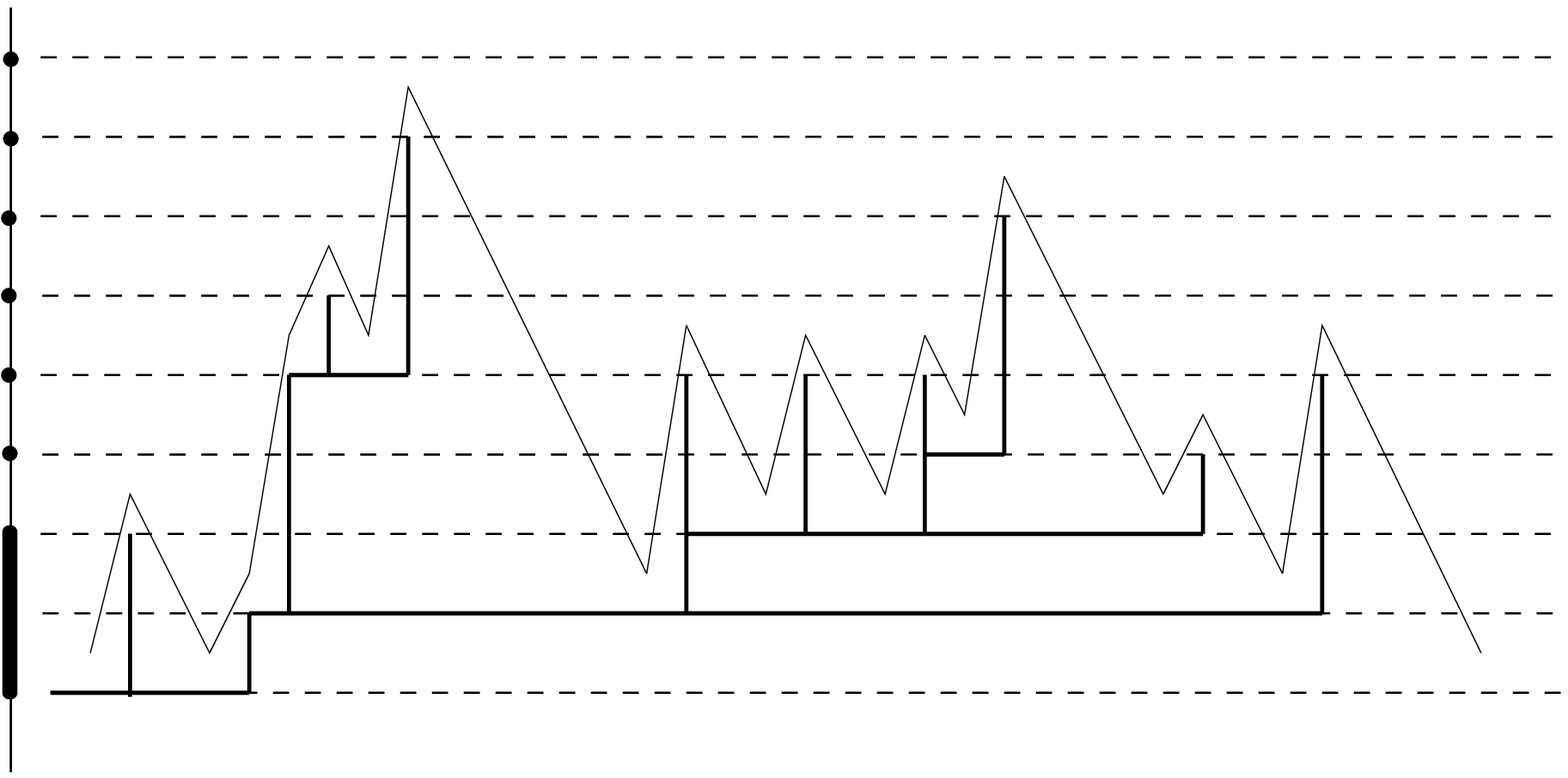}{12.truecm}
\figlabel\multirw

As before, these configurations are in one-to-one correspondence
with discrete Random Walks from $0$ to $0$ on the integral half-line $[0,\infty)$,
with possible ascending steps of $+i$, $i=1,2,...,k$, weighted by 
$t_i$, and descending steps $-1$
(see Fig.\multirw). The partition function $\mu$ of these walks clearly 
satisfies the self-consistent equation \charmut\ corresponding
to a decomposition of the walks according to the length $i$ of their first
ascending step, with $\mu^{i+1}$ 
being the partition function of the rest of the walk going from position $i$ back 
to the origin on the half-line.

Another quantity of interest is the so-called loop-loop propagator
$Z_{T}(i,j\vert \{t_m\})$
defined as the partition function of surfaces with $T$ time slices
and with fixed numbers $i$ (resp. $j$) of tiles originating  
(resp. terminating) in the first (resp. last) slice.
The inversion formula \fundafin\ extends straightforwardly to include an extra weight
$w_s$ attached to each portion of tile visiting the slice $s=0,1,...,T$.  
Choosing $w_0=x$, $w_T=y$ and $w_s=1$ for $s=1,2,...,T-1$, we immediately get the
generating function $G_T(x,y\vert \{t_m\})=\sum_{i,j\geq 0} x^i y^j Z_{T}(i,j\vert
\{t_m\})$ as the inverse of the partition function 
$G_T^{hm}(x,y\vert \{z_m\})$ of hard multimers with
weights $z_m=-t_m$ for $(m+1)$-mers and extra weights $x$ (resp. $y$) for the
multimers originating (resp. terminating) 
at point $0$ (resp. $T$).
To compute $G_T^{hm}(x,y\vert \{z_m\})$ 
we make the simple observation that it is at most
linear in the variables $x$ and $y$ as at most one multimer can touch the point $0$
(resp. $T$), and that it is symmetric in $x$ and $y$. Writing
\eqn\abc{ G_T^{hm}(x,y\vert \{z_m\})= a -b (x+y) +c xy }
we easily compute the coefficients $a,b,c$ by taking particular values of $x$ and $y$.
For $x=y=0$, we forbid multimers to touch the extremities of the segment $[0,T]$
thus effectively reducing it to $[1,T-1]$ and resulting in $a=Z_{T-2}^{hm}(
\{z_m\})$. For $x=1$ and $y=0$ we only forbid the $T$ point, with the result
$a-b=Z_{T-1}^{hm}(\{z_m\})$. Finally, for $x=y=1$ we simply have 
$a-2 b+c=Z_{T}^{hm}(\{z_m\})$.
This results in the following expression for the loop-loop propagator
\eqn\loplop{ G_T(x,y\vert \{t_m\})={1\over Z_{T-2}^{hm}-(x+y)
(Z_{T-2}^{hm}-Z_{T-1}^{hm}) +x y (Z_{T-2}^{hm}-2Z_{T-1}^{hm}+Z_{T}^{hm})} }
where the $Z_{N}^{hm}$ are all taken at the values $z_m=-t_m$.
Explicitly expanding this as a series in $x$ and $y$, we obtain
\eqn\loope{ Z_{T}(i,j\vert \{t_m\})= {1\over Z_{T-2}^{hm}} \left(1-{Z_{T-1}^{hm}\over
Z_{T-2}^{hm}}\right)^{i+j} \sum_{r\geq 0} {i\choose r} {j \choose r} \left(
{(Z_{T-1}^{hm})^2-Z_{T}^{hm} Z_{T-2}^{hm}\over (Z_{T-2}^{hm}-Z_{T-1}^{hm})^2} \right)^r}

\subsec{ First application: hard $(k+1)$-mers and Fuss-Catalan numbers}

As a first simple application of our construction, let us consider
surfaces made only of time-like $2(k+1)$-gons, i.e. with duals made
only of vertical edges of fixed length $k$. 
This corresponds to specializing
the above to $t_i=\delta_{i,k} t$, where $t$ is the weight per tile.
In the limit of large $T$, this yields
the thermodynamic partition function 
$Z_T^{(k)}(t) \sim \mu^T(t)$ where $\mu(t)$ is the smallest solution
(in module) of $t\mu^{k+1}=\mu-1$, known as the generating function of the
Fuss-Catalan numbers $c_n^{(k)}$ \BIJO\ 
\eqn\fussca{\mu(t)=\sum_{n\geq 0} c_n^{(k)} t^n \ \ {\rm where} \ \ 
c_n^{(k)}={((k+1)n)! \over (kn+1)! n!} }
{}From the use of Stirling's formula, we get the large $n$ behavior
of  these numbers: $c_n^{(k)}\sim \big((k+1)^{k+1}/k^k\big)^n/n^{3/2}$,
which allows to show that
for all $k\geq 1$, the function $\mu(t)$ displays a square root singularity 
$\mu\vert_{sing}\sim \sqrt{t_c-t}$
when $t$ approaches the critical value
\eqn\critifusca{ t_c= {k^k \over (k+1)^{k+1}} }
Therefore the scaling limit of these models (for all $k\geq 1$) 
lies in the same universality class
as that of pure Lorentzian triangulations, corresponding to $k=1$.

Note finally that according
to the above equivalence with Random Walks, the function $\mu(t)$ also generates
the numbers of directed Random Walks on the integer half-line $[0,\infty)$ 
starting and ending at $0$ and with ascending steps of $+k$ only and descending
steps of $-1$, and with a weight $t$ per ascending step.

\subsec{Second application: multicritical models of $(1+1)$-D Lorentzian surfaces
and Patalan numbers}

In this section we show how to go beyond the generic square root singularity
of pure Lorentzian surfaces and get more interesting critical behaviors.  
As usual, this can be done by fine-tuning the weights $t_i$ in order to reach
multicritical points. Indeed, we may reach a multicritical point of
order $(k+1)$ by retaining di-, tri-, ... , $(k+1)$-mers and fine-tuning
the activities $t_i$, $i=1,2,...,k$ in order for eqn. 
\charmut\ to take the form
\eqn\formut{ \big(1-{2\over k} t\mu\big)^{k+1} =1 -{2(k+1)\over k} t }
namely by picking
\eqn\pick{ t_i = -{1\over k+1} {k+1 \choose i+1} \left(- {2 t\over k} \right)^i\ ,
\  \ i=1,2,... k }
all expressed in terms of the activity $t=t_1$ per lozenge.
The values of the coefficients in \formut\ are entirely fixed by the constant, linear
and quadratic terms in \charmut, with relative values $1$, $-1$ and $t$. 
This yields the thermodynamic partition function per time slice
\eqn\mumulti{ \mu(t) = {k \over 2 t}
\left(1 - \left(1-{2(k+1)\over k} t\right)^{1\over k+1}\right)}
For instance, for $k=2$, we reach 
a tricritical point by 
taking activities $t$ per lozenge and $-t^2/3$ per hexagon, 
leading to $\mu=(1-(1-3t)^{1\over 3})/t$.
The need for both positive and negative activities to reach a multicritical
point parallels the case of ordinary random surfaces as solved by means
of the one-matrix models, whose potentials display the same pattern of
alternating signs for the activities per tiles.
Note also that similarly to that case, we need to consider {\it at least}
$(k+1)$-mers to reach a multicritical point of order $(k+1)$. In that respect,
eqn.\formut\ is the minimal realization of the $(k+1)$-critical point in which
the requirement of multicriticality fixes 
all the $t_i$'s in terms of $t=t_1$. The same point could be attained from any
model involving also larger multimers leading to a characteristic equation of degree
larger than $k+1$. The corresponding $(k+1)$-critical point involves then $k-1$
relations between $t_1,t_2,...,t_k,...$.

Up to the change of variables $t=k(k+1) x/2$, the function $\mu=\sum_{n\geq 0}
p_n^{(k)} x^n$
\mumulti\  is known as the 
generating function of so-called  Patalan numbers \PATA\ 
\eqn\pata{ p_n^{(k)} = {(k+1)^n \over (n+1)!} \prod_{i=1}^n ((k+1)i-1) }
which are all positive integers. The integrality of these numbers
is clear from eqn. \pick, since all the $t_i$ are integer multiples of $x^i$.
On the other hand, the positivity appears as non-trivial, as
the $t_i$ have alternating signs. This strongly suggests a possible
purely combinatorial reinterpretation of our multicritical partition functions.

The singularity of \mumulti\ leads to a new scaling behavior for the
corresponding decorated surfaces, when $t$ approaches the
multicritical point $t_c=k/(2(k+1))$. Indeed, to get a proper scaling limit
of the multicritical partition function $Z_T^{[k]}(t)=Z_T(\{t_i\})\sim \mu^T$,
corresponding to the fine-tuning \pick\ 
and with $\mu$ as in \mumulti, 
we must set
\eqn\setprop{ T={\tau \over a} \qquad t=t_c(1- a^{k+1} \Lambda) }
where $\tau$ is the renormalized time lapse and
where $\Lambda$ is the renormalized activity per unit area of tile
(``cosmological constant"), with $a\to 0$ in the scaling regime.    
In this regime, $(\mu(t)/\mu(t_c))^T \to e^{-\tau \Lambda^{1\over k+1}}$ 
and more generally 
$Z_T^{[k]}(t)/Z_T^{[k]}(t_c)$ becomes a universal scaling 
function of $\tau \Lambda^{1\over k+1}$ which we will 
determine below\foot{Note that this scaling function involves
$\mu$ and also the other roots of the
polynomial equation \formut\ as they all merge at the multicritical point.}.  
Introducing the area $A=\sum_{i\geq 1} i N_i$ where $N_i$ is
the total number of $2(i+1)$-gonal tiles forming the surface,
we see that the total  Boltzmann weight is proportional to $t^A$.
Writing $t^A\sim t_c^A e^{-A{t_c-t\over t_c}}$ when $t\to t_c$,
and substituting \setprop, we see that  
the parameter $\Lambda$ is conjugate
to the renormalized area of the surface, defined as ${\cal A}=a^{k+1}A$. 
We immediately deduce that ${\cal A}\sim \tau^{k+1}$
and get the scaling behavior of the
area of the surface 
\eqn\scarea{A\sim T^{k+1}}
in terms of the time lapse $T$ when $T$ is large. 
This determines the ``fractal dimension of space-time" to be $d_F=k+1$. 
In the directed Random Walk picture of previous section, 
this allows to obtain the scaling
behavior of the vertical extent or
gyration radius $R=T$ of the walks on $[0,\infty)$, 
in terms of their total length
$L=2A$, namely $R\sim L^\nu$ with the exponent
\eqn\exporw{ \nu={1\over d_F}={1\over k+1} }
This is also the exponent of the correlation length $\xi\sim (t_c-t)^{-\nu}$
appearing for instance in the correlation of the numbers of tiles
${\cal N}(T_1), {\cal N}(T_2)$ in two
given time slices at times $T_1,T_2$, namely $\langle {\cal N}(T_1){\cal
N}(T_2)\rangle \sim e^{-{|T_1-T_2|\over \xi}}$.  
The partition function of the walks of fixed length $L$ is identified with
the coefficient of $t^{L\over 2}$ in the expansion of $\mu(t)$ \mumulti,
and behaves as
$(t_c)^{-{L\over 2}} L^{\alpha-3}$ with the susceptibility exponent
\eqn\alval{ \alpha= {2k+1 \over k+1} }
Equivalently, in the language of random surfaces, the coefficient of 
$t^A$ in the expansion of $\mu(t)$ \mumulti\ represents the 
thermodynamic partition
function $Z_A$ of fine-tuned semi-random surfaces of fixed area $A$
and with fixed projection either empty
or equal to $C_0$, namely reduced to either zero or a single 
dimer. Therefore
the exponent $\alpha$ \alval\ is interpreted as a configuration exponent
for these objects, namely $Z_A\sim (t_c)^{-A} A^{\alpha-3}$ for large $A$.
Note finally that, as expected, the exponents $\alpha$ and $\nu$ 
above obey the hyperscaling relation
$2-\alpha =d \nu$ in $d=1$ dimension.
 
To conclude this section, let us derive the explicit form of 
the finite-time partition function $Z_T^{[k]}(t)$ for the
fine-tuned semi-random surfaces and its scaling limit, together
with that of the corresponding loop-loop propagator. 
Starting from the inversion formula 
\fundafin, we are left with the calculation of the partition function of 
fine-tuned hard multimers $Z_T^{hm[k]}$ expressed as
\eqn\tmathm{ Z_T^{hm[k]}= v^t \big({\cal T}_k\big)^{T+1} v }
where the $(k+1)\times (k+1)$
transfer matrix reads as in \transinf\ with $z_i=-t_i$ given by \pick,
and where $v^t=(1,0,0,...,0)$. The eigenvalues $\lambda_j$,
$j=0,1,...,k$ of this transfer matrix are simply 
the inverses of the solutions $\mu_j$
to the multicritical characteristic equation \formut, reading
\eqn\redsol{ \mu_j= {k \over 2 t} (1-\omega^j\Delta) \qquad {\rm with} \ \ 
\omega=e^{2{\rm i}{\pi \over k+1}} \ , \ \ \ \Delta=\big(
1-{2(k+1)\over k}t\big)^{1\over k+1}}
The partition function \tmathm\ is expressed as a linear combination
$Z_T^{hm[k]}=\sum_{0 \leq j \leq k} a_j/(\mu_j)^{T+1}$, where the coefficients
$a_j$ may be obtained by explicit diagonalization of ${\cal T}_k$.
After some algebra we find the rather simple expression
\eqn\simple{ Z_T^{hm[k]}={1\over \Delta^k} 
\left( {2t\over k} \right)^{T+2} \sum_{j=0}^k {\omega^j \over (1-\omega^j
\Delta)^{T+2} }}
which reduces to \hdonepart\ for $k=1$. Note that it satisfies the 
recursion relation $Z_{T+1}^{hm[k]}=Z_T^{hm[k]}-\sum_{i=1}^k
t_i Z_{T-i}^{hm[k]}$ with the fine-tuned values of $t_i$ \pick. The solution
\simple\ is the unique function obeying this recursion relation with the
initial conditions: $Z_{-i}^{hm[k]}=0$, $i=2,3,...,k+1$ and $Z_{-1}^{hm[k]}=1$. 
For illustration,
let us display the first few values of $Z_T^{hm[k]}$ for non-negative $T$:
\eqn\dispzt{\eqalign{
Z_0^{hm[k]} &= 1\cr
Z_1^{hm[k]} &= 1-t \cr
Z_2^{hm[k]} &= 1-2t+{2(k-1)\over 3 k} t^2 \cr
Z_3^{hm[k]} &= 1-3t+{7k-4\over 3 k} t^2
-{(k-1)(k-2)\over 3 k^2}t^3 \cr}}
By use of the inversion formula \fundafin,
the expression \simple\ finally leads to the partition function of multicritical 
semi-random surfaces
\eqn\parlutfin{ Z_T^{[k]}(t)= \left( {k \over 2t} \right)^{T+2} { \Delta^k \over
\sum_{j=0}^k {\omega^j \over (1-\omega^j
\Delta)^{T+2} } }}
Note that as $t\to t_c=k/(2(k+1))$, $\Delta\to 0$, the partition function
function \parlutfin\ tends to a finite limit, as the denominator is of
order $\Delta^k$. This yields a finite ratio
\eqn\ratfinite{
{Z_T^{[k]}(t)\over Z_T^{[k]}(t_c)}= (k+1)\left({t_c\over t}\right)^{T+2}
{T+1+k \choose k}
{\Delta^k \over \sum_{j=0}^k {\omega^j \over (1-\omega^j \Delta)^{T+2} }}}
We may now perform the scaling limit \setprop\ of this
ratio, that tends to a universal scaling function of the variable
$x=\tau \Lambda^{1\over k+1}$, namely 
\eqn\unirat{ \lim_{a\to 0}{Z_T^{[k]}(t)\over Z_T^{[k]}(t_c)}=
{k+1 \over k!} {x^k \over
h_1(x)}, \qquad x=\tau \Lambda^{1\over k+1}, }
where we have used the generalized hyperbolic
function $h_1$ of order $k+1$, member of the family  
$h_p(x)$, $p=0,1,2...,k$, defined by
\eqn\hyperbol{ h_p(x) = \sum_{j=0}^k \omega^{pj} e^{ \omega^j x} }
with $\omega$ as in \redsol.
The functions $h_p$ are cousins of the Mittag-Leffler function 
\BAT, generalizing
the hyperbolic sine and cosine (case $k=1$, $p=1,2$ respectively).
For $k=1$ indeed, with $h_1(x)=2\sinh(x)$, \unirat\ allows to
recover the usual scaling function $x/\sinh(x)$ \FGK. 

We can fix the normalization of the partition function in order for its
scaling limit to read
\eqn\newz{ {\cal Z}_\tau^{[k]}(\Lambda)\equiv {1\over (k+1)!} \lim_{a\to 0} 
{Z_T^{[k]}(t)\over (k+1)^{T}}=
{(k+1) \Lambda^{k \over k+1} \over 
k! h_1(\tau \Lambda^{1\over k+1})} }
and accordingly the critical value ${\cal Z}_\tau^{[k]}(0)=1/\tau^k$.
In particular, this allows to recover for $k=1$ the
partition function 
${\cal Z}_\tau^{[1]}(\Lambda)=\sqrt{\Lambda}/\sinh(\tau\sqrt{\Lambda})$ 
of \FGK.

Let us now turn to the loop-loop propagator \loope. To define a sensible 
scaling limit of this quantity we need to take
\eqn\moresca{ i={L_1 \over a} \qquad j={L_2 \over a} }
and to substitute in \loope\ the value of the hard multimer partition function
\eqn\lartzt{ Z_{T-2}^{hm}\propto {1\over (k+1)^T} h_1(\tau \Lambda^{1\over k+1}) }
with $h_1$ as in \hyperbol. Indeed, 
except for the prefactor $1/Z_{T-2}^{hm}$ which we normalize as in \newz, 
only ratios of $Z_T^{hm}$'s enter the
expression \loope\ so we may forget about all the prefactors independent of $T$
as indicated in \lartzt\ by the proportionality symbol. 
Eqn. \loope\ finally takes the  scaling form
\eqn\newloop{ G_\tau^{[k]}(L_1,L_2) = {\cal Z}_\tau^{[k]}(\Lambda)  
\ e^{-(L_1+L_2) {\Lambda^{1\over k+1} \over k} {h_2\over
h_1} } \ I_0\left( 2\sqrt{L_1L_2 {\Lambda^{2\over k+1}\over
k^2} { h_2^2 - h_1 h_3 \over h_1^2}}\right) }
with ${\cal Z}_\tau^{[k]}(\Lambda)$ as in \newz, with 
$I_0(2x)=\sum_{r\geq 0}x^{2r}/(r!)^2$
the modified Bessel function, and
where the hyperbolic functions $h_p\equiv h_p(\tau \Lambda^{1\over k+1})$
are defined in \hyperbol.
For completeness,
let us display the hyperbolic functions entering the loop-loop propagator for 
the first few values of $k=1,2,3$.
\eqn\firstfew{ \eqalign{ k=1:&\ \  h_1(x)=h_3(x)=2 \sinh(x) \qquad h_2(x)=2 \cosh(x)\cr
k=2:&\ \  h_m(x)=e^x +2 e^{-{x\over 2}} \cos({\sqrt{3}\over 2}x +2m {\pi\over 3})
\ m=1,2,3 \cr 
k=3:&\ \  h_1(x)=2 (\sinh(x)-\sin(x)) \quad h_2(x)=2(\cosh(x)-\cos(x)) \cr
&\ \ h_3(x)=2(\sinh(x)+\sin(x)) \cr}}
With these, we recover in the $k=1$ case the results of \FGK.

Let us finally comment on the form of the rescaled partition function
${\cal Z}_\tau^{[k]}(\Lambda)$ \newz. 
In terms of the rescaled
variable $x=\tau \Lambda^{1\over k+1}$, it has poles situated at the complex 
non-vanishing zeros of $h_1(x)$, all
of the form $x_{j,m}=-\omega^j\alpha_m^{1\over
k+1}$, $j=0,1,2...,k$ and $m=1,2,3,...$ with $\alpha_m$ real positive. 
This allows to write the partition function as
\eqn\pfprod{ {\cal Z}_\tau^{[k]}(\Lambda) 
= {1\over \tau^{k+1} \prod_{m=1}^\infty
(1+{\Lambda \tau^{k+1} \over \alpha_m}) } }
When $k=1$ we simply have $\alpha_m = \pi^2 m^2$, $m=1,2,3...$. For arbitrary $k$
there is no such simple form, but asymptotically one can show that
\eqn\asymzer{ \alpha_m ={\pi^{k+1}\over \sin^{k+1}({\pi\over k+1})} 
\left(m+{k-1 \over 2(k+1)} +O(e^{-\epsilon m}) \right)^{k+1} }

\newsec{$(1+2)$-D Lorentzian tetrahedral complexes and
the hard hexagon model}

In this section, we consider three-dimensional
random objects made of tetrahedra, corresponding to $(1+2)$-D
Lorentzian tetrahedral complexes, 
with two (time-like) regular directions and one 
(space-like) random one. This is to be contrasted with the more complicated
situation of discrete $(2+1)$-D Lorentzian manifolds used in the context
of $3$-D gravity \AMBJ, which have one (time-like) regular direction 
and two (space-like)
random ones. Still, by slightly breaking the symmetry in the two time
directions, we shall be able to view the configurations
of our model as the time-evolution of some particular  
random triangulations, namely the Lorentzian ones.

\subsec{The plaquette model for Lorentzian tetrahedral complexes}

\fig{A sample hexagonal plaquette configuration (a). There are arbitrarily
many plaquettes in each tube. Neighboring plaquettes cannot
cross each other, neither within the same tube, nor between
adjacent tubes. Such configurations are dual to 
tetrahedral complexes, made of tetrahedra filling triangular tubes
dual to the hexagonal ones. We have represented an elementary 
diamond-shaped building block
(b) dual to a plaquette of (a), and made of six tetrahedra sharing
an edge.}{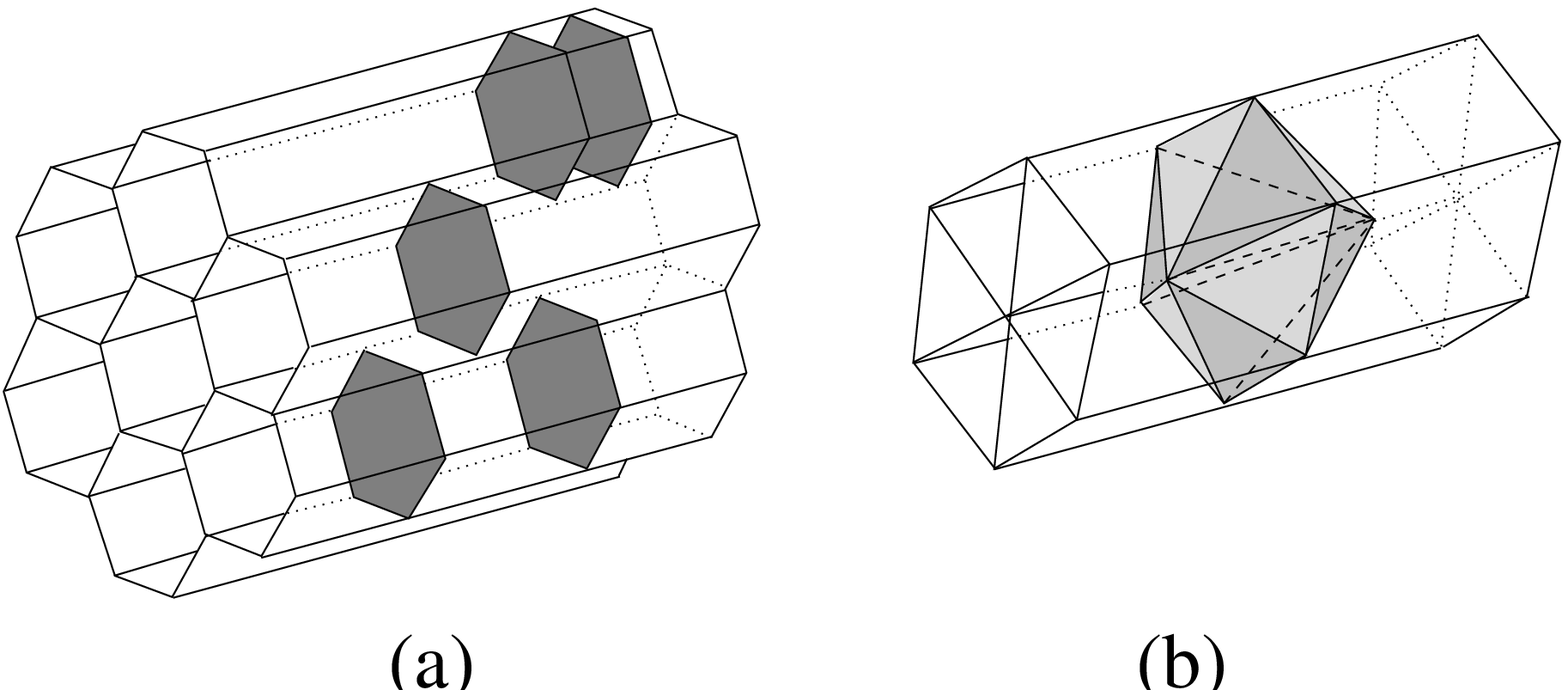}{12.cm}
\figlabel\deuxD

If we view the pure $(1+1)$-D Lorentzian triangulations in the dual picture
as a regular array of slices in which edges may freely slide horizontally
provided they do not cross one-another, a natural $(1+2)$-D generalization
consists in having a regular $2$D array of horizontal tubes in which plaquettes
may freely slide horizontally, provided they do not cross one-another.
More precisely, we consider here the situation depicted in
Fig.\deuxD\ (a), 
where the tubes have a regular hexagonal section, and their 
array forms the two-dimensional hexagonal lattice. The plaquettes are
filled hexagons orthogonal to the tubes and with the same section. 
A configuration is characterized by an arbitrary arrangement of such plaquettes
in the tubes, where we take into account the relative ordering of the 
plaquettes with respect to one-another. In practice, a given plaquette 
only sees its nearest neighbors within the same tube and those in the
six adjacent tubes, through their edges.
For definiteness we attach a weight $t$ per plaquette and consider only a
finite array of tubes with transverse size $T_1\times T_2$ 
and with a bulky shape.
With this definition, we generate three-dimensional
objects with two (time-like) regular directions and one 
(space-like) random one.

Like in the case of triangulations where the edges are
dual to pairs of triangles forming lozenges, 
the above hexagonal plaquettes are dual to a diamond-shaped
dodecahedral volume made of $6$ tetrahedra glued around a common edge, as
shown in Fig.\deuxD\ (b). 
The role formerly played by triangles is now played by tetrahedra:
these live inside tubes with triangular section (dual to the original
hexagonal tubes). Each such triangular tube is tessellated by 
tetrahedra of three possible kinds: each tetrahedron has $1$ vertex
on each edge of the tube, and the three kinds correspond to the 
three possible positions of the fourth vertex on one of the three
edges of the tube\foot{Note that this generalizes 
straightforwardly the $(1+1)$-D situation, where
the triangles in each slice are of two kinds: those pointing up 
and those pointing down.}. 
Therefore exactly one edge of the tube contains one 
(space-like) edge of the tetrahedron.
This edge is also common to the five other tetrahedra which together
with the above form a diamond-shaped building block.
All tetrahedra in a tube are glued along their triangular faces 
to two others in the same tube and two others in the two 
adjacent tubes sharing  the abovementioned space-like edge. 

\fig{The regular tetrahedral complex is nothing but the FCC
lattice, made of octahedra supplemented by tetrahedra as shown.
To get only tetrahedra, we simply add a diagonal edge inside all 
octahedra as indicated. This allows to decompose each octahedron into
four tetrahedra. The building block of our model is composed
of $6$ tetrahedra, $2$ from the original FCC lattice, sharing one edge,
and $2$ from each decomposed adjacent octahedron.}{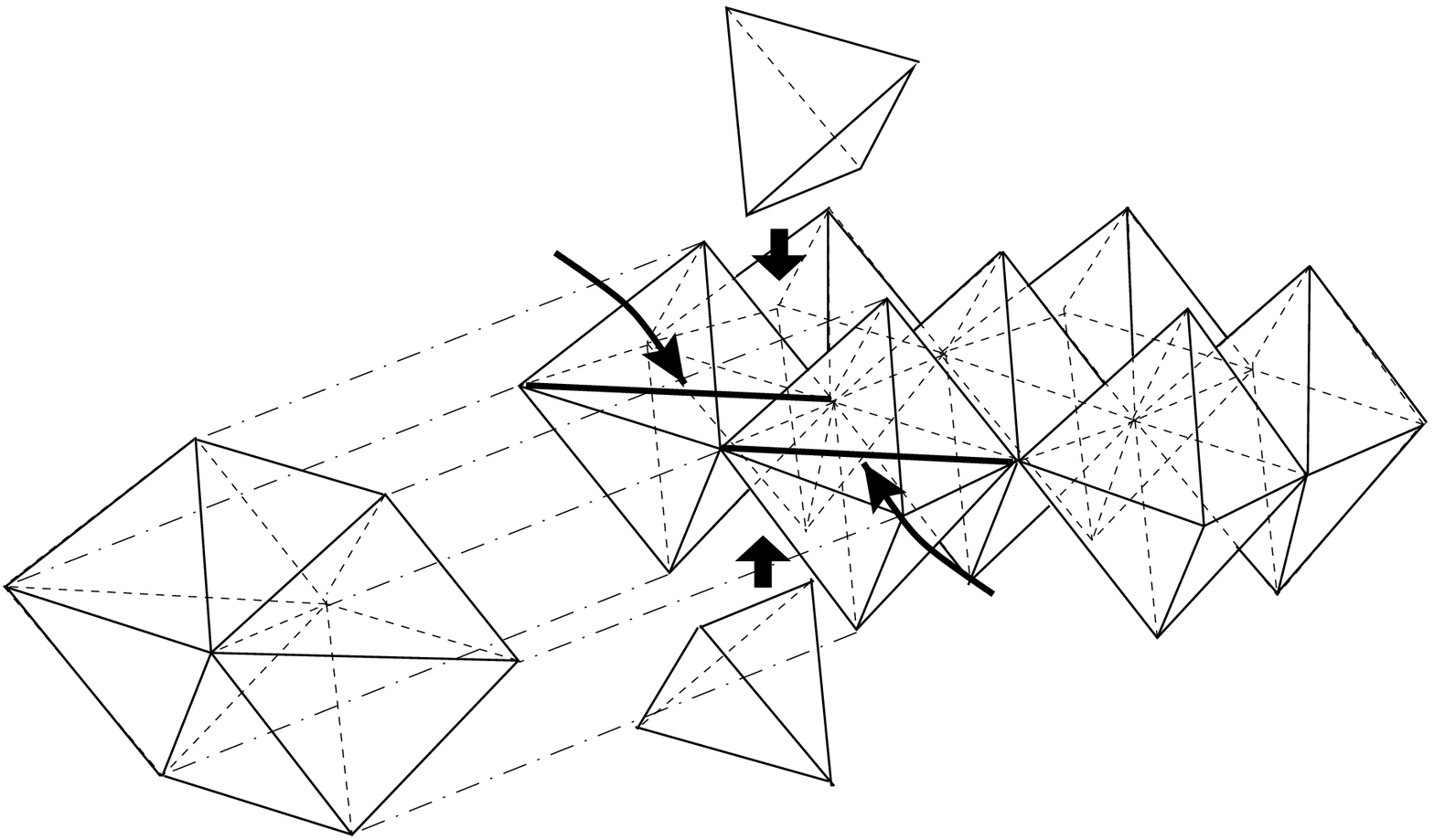}{10.cm}
\figlabel\FCC

It is interesting to view our tetrahedral complexes as a random
version of some regular lattice, in the same way as the Lorentzian
triangulations are deformations of the triangular lattice.
To construct a regular lattice, we simply need to fill the
triangular tubes with tetrahedra in a regular way, say by taking
a succession of tetrahedra of the first, second, third, first, etc... kinds
along each tube. The other choice of chirality (first, third, second, first,
etc...) is completely equivalent.
The resulting lattice is nothing but a FCC lattice (see Fig.\FCC).  
Indeed, the latter can be viewed as a regular arrangement of octahedra
completed by tetrahedra. Each octahedron can be decomposed into
four tetrahedra by adding a diagonal edge (see Fig.\FCC). 
One possible elementary cell is obtained by first considering  
two original tetrahedra of the FCC lattice sharing one edge, and then by
completing them by the four adjacent tetrahedra (sharing the same edge),
two from each neighboring octahedron.  
These $6$ tetrahedra form the diamond-shaped building block of our model.
The two choices of chirality  correspond to the two possible
choices of diagonal edge added inside the octahedra. 
In the dual plaquette language, the FCC lattice corresponds to the 
following regular arrangement of plaquettes: we decompose the 
(triangular) lattice of hexagonal tubes into three sublattices
denoted, say $1$, $2$, $3$, such that no
two adjacent tubes belong to the same sublattice. Plaquettes are 
then arranged in equidistant successive parallel planes in such a
way that plaquettes in successive planes cover the three sublattices
$1$, $2$, $3$, $1$, ...
alternatively. 

\fig{The horizontal section of a sample $(1+2)$-D tetrahedral complex
with vertical size $T_2$ and horizontal size $T_1$ is a
$(1+1)$-D Lorentzian triangulation over a time lapse 
$T_1$.}{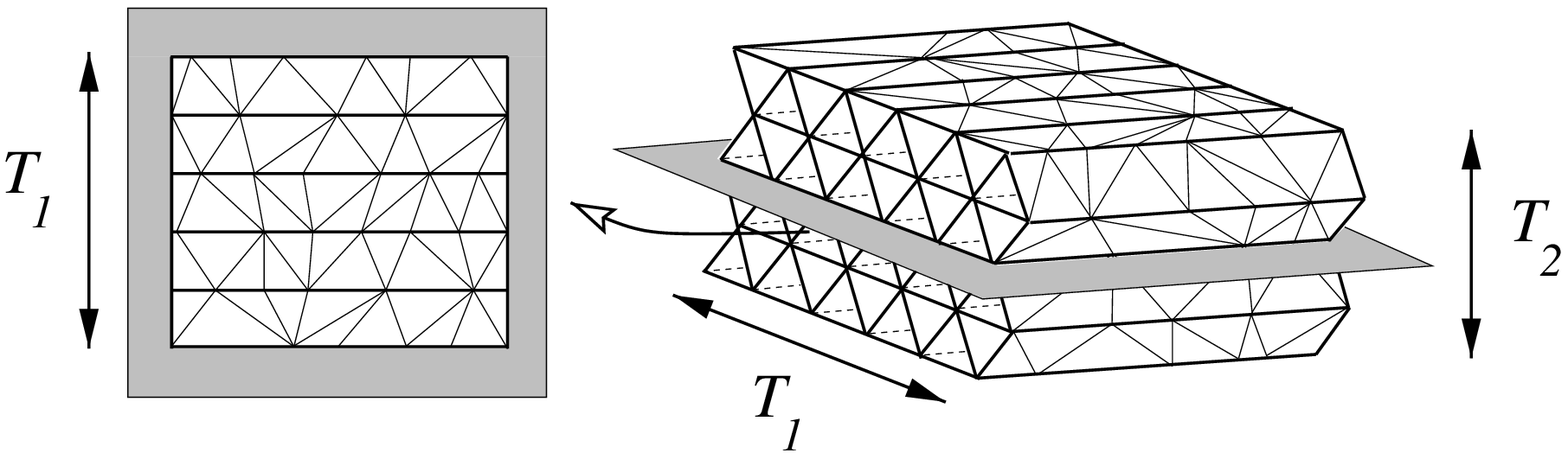}{13.cm}
\figlabel\briques

The restriction of any given  tetrahedral complex to a plane generated
by the longitudinal direction of the tubes and one of the three directions
of the triangular lattice section of the tubes is nothing but 
a particular $(1+1)$-D Lorentzian triangulation (see Fig.\briques).
We may therefore view the configurations of our model as the
time evolution of such triangulations in successive parallel planes (with
a total of $T_2$ plane slices), each of which extends over a fixed
time lapse $T_1$ as shown in Fig.\briques. 
The constant time-lines for the Lorentzian triangulations
in two successive planes are shifted by half a period of the triangular
lattice of tubes: indeed, these are nothing but edges of a same triangular
tube. Each triangle in a given plane is linked to a vertex
of the triangulation of the next plane, belonging to the same tube. 
These ``face-to-vertex" tetrahedra clearly
coexist with ``vertex-to-face" ones, but also with ``edge-to-edge" ones,
namely with exactly one edge in each plane (one space-like and one 
time-like), and only time-like triangular faces. 

\fig{The dual configurations of two consecutive plane sections
of a sample $(1+2)$-D Lorentzian tetrahedral complex. The lower
(resp. upper) configuration is represented by dashed (resp. solid) lines.
The constant time lines of the two configurations are shifted by half
a period. We have encircled the three types of vertices of the resulting
mixed graph, in one-to-one correspondence with the
three types of tetrahedra filling the slice between the two planes
(see text).}{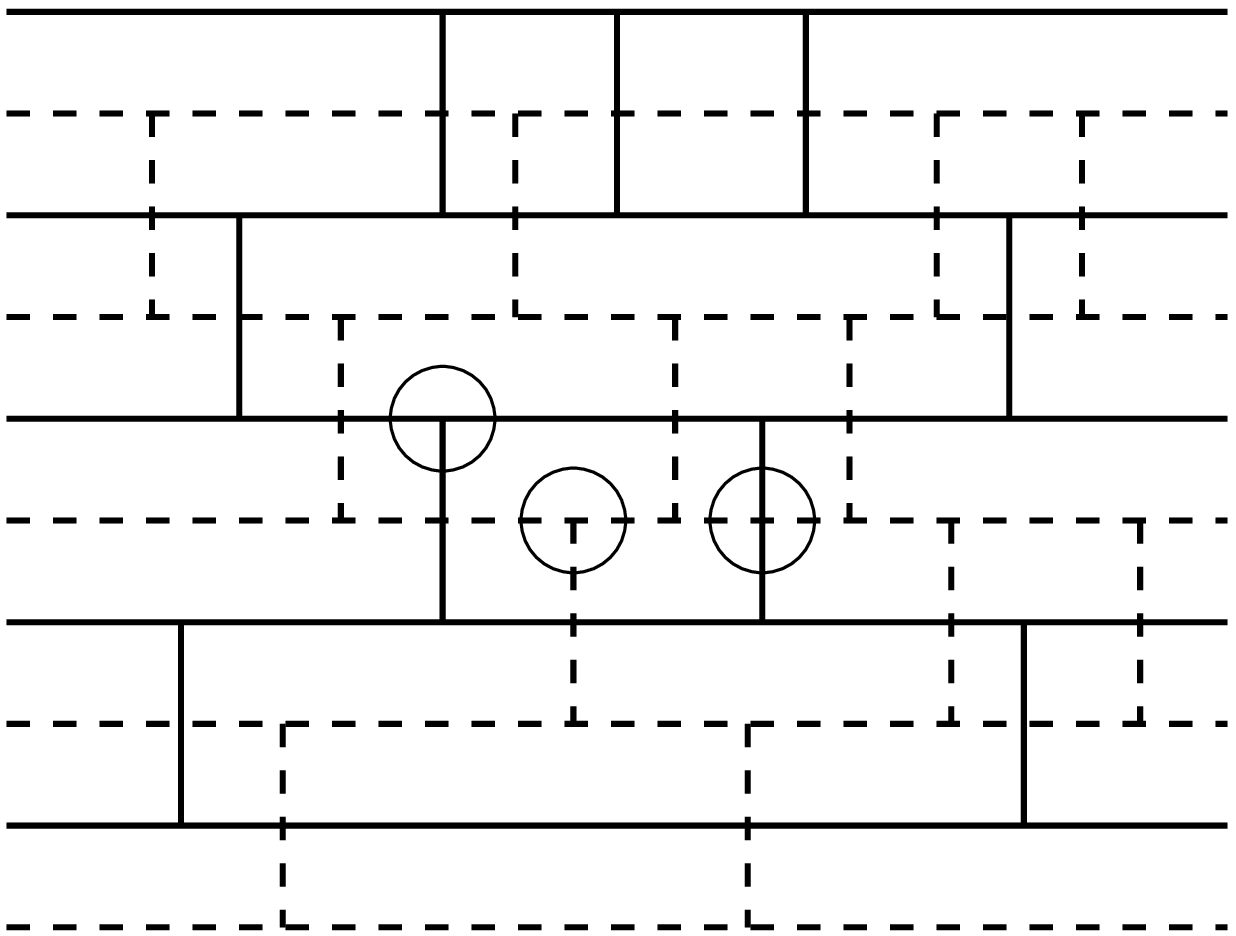}{8.cm}
\figlabel\intersec

The dual picture allows for a better understanding of the tetrahedral
complex 
between two consecutive planes \AMBJ. Indeed, the dual configurations
of the triangulations in the  lower and upper planes
may be represented  as in  Fig.\intersec\
in dashed and solid lines on the same picture, with their
time-lines shifted by a half-period. The three types of vertices
(solid trivalent, dashed trivalent and mixed tetravalent)
of the resulting mixed graph are in one-to-one
correspondence with the three types of tetrahedra 
(face-to-vertex, vertex-to-face and edge-to-edge)
in the slice between the two planes.

\subsec{Equivalence with hard hexagons and critical behavior}

We now wish to evaluate the partition function $Z_{\cal D}(t)$ for
the Lorentzian tetrahedral complexes in the plaquette formulation 
inside hexagonal tubes with a total section made of a compact 
connected domain $\cal D$ of the hexagonal lattice. 
For each plaquette configuration, we define a left projection
obtained by letting the leftmost plaquettes slide along the tubes all the
way to their left end. The projections thus obtained are in 
one-to-one correspondence with configurations of hard hexagons on the 
two-dimensional domain $\cal D$.  By hard hexagons, we mean that 
if an hexagon of $\cal D$ is occupied, then its immediate neighbors
must be empty\foot{In the standard picture, the hardness constraint
is implemented by a weaker {\it no-overlap} constraint for slightly larger
hexagons, obtained from our hexagonal plaquettes by rotating them by
$\pi/6$ and dilating them by $\sqrt{3}$. These larger hexagons
are also the projections of our diamond-shaped building blocks. The two ways
of implementing the hardness constraint are clearly equivalent.}.
Repeating the arguments of Section 2.1, we arrive at the following inversion
formula:
\eqn\formid{ Z_{\cal D}(t)={1\over Z_{\cal D}^{hh}(-t)} }
where we have denoted by $Z_{\cal D}^{hh}(z)$ the partition function of
the hard hexagon model on the domain $\cal D$ with an 
activity $z$ per occupied hexagon.
We also have the more general formulas generalizing \corr:
\eqn\formicorr{
Z_{\cal D}^{(C)}(t)= {Z^{hh}_{{\cal D}\setminus C}(-t)\over 
Z_{\cal D}^{hh}(-t)} }
where $Z_{\cal D}^{(C)}(t)$ is the partition function for the configurations
with left projection $C\subset {\cal D}$ and with the weight of the projection
removed, while  
$Z^{hh}_{{\cal D}\setminus C}(z)$ is the hard hexagon partition function
on the domain ${\cal D}\setminus C$, i.e. the set of hexagonal tubes in
${\cal D}$ with no common edge with hexagons of $C$.
The latter quantities \formicorr\ have a well-defined thermodynamic 
limit, obtained by keeping $C$ fixed, and
letting the domain $\cal D$ become infinite.
In particular, if $C$ is reduced to a single plaquette $C_0$, the 
quantity $Z_{\cal D}^{(C_0)}(t)$ tends to the thermodynamic
density of occupied hexagons $\rho_{hh}(z=-t)$ of the hard hexagon model.
Note that this partition function $Z_{\cal D}^{(C_0)}(t)$ is the
generating function of the number $N_V$ of tetrahedral complexes with $V$
plaquettes (i.e. tetrahedral complexes 
with  volume $V$ if we assign a unit volume to the
diamond-shaped building block)
and projection a single one $C_0$. Note that the constraint
that the projection be reduced to a single plaquette amounts to generalized
``staircase" boundary conditions. 

We may now use Baxter's exact solution of the
hard hexagon model \BAXHH\ to obtain the critical 
behavior of $Z_{\cal D}^{(C_0)}(t)$.
The hard hexagon is known to have two critical singularities 
at a positive and a negative values of $z$. 
The singularity of our problem
occurs at the negative value $z_c=\big({1-\sqrt{5}\over 2}\big)^5
={11-5\sqrt{5}\over 2}$ of $z$ 
(corresponding to a positive $t_c=-z_c$), where the model is known to belong
to the universality class of the Lee--Yang edge singularity \KF\ \DSZ,
itself described by a conformal field theory of central charge $c=-22/5$
\CARTWO.
The thermodynamic partition function per
hexagon $\kappa_{hh}(z)$ behaves as $z\to z_c$ as
\eqn\kapbeh{ \kappa_{hh}(z)= \kappa_0 + \kappa_1 (z-z_c)^{5\over 6} +O(z-z_c)}
corresponding to the critical exponent $\alpha=2-5/6=7/6$,
and we also have
\eqn\densit{ \rho_{hh}(z)= {d \over dz} {\rm Log}\, \kappa_{hh}(z)\sim {5\over
6}{\kappa_1\over \kappa_0} (z-z_c)^{-{1\over 6}} } 
{}From this result, we immediately deduce that 
$Z_{\cal D}^{(C_0)}(t)\sim(t_c-t)^{-1/6}$, therefore 
the number $N_V$ of tetrahedral complexes with $V$ plaquettes
and projection a single one behaves for large $V$ as
\eqn\behaV{ N_V \sim \ {\rm const.} {\left({11+5\sqrt{5} \over 2}\right)^V
\over V^{5\over 6} } }
Returning to the case of a finite but large domain $\cal D$ with area
$T_1T_2$, we may write the partition function $Z_{\cal D}(t)$
at leading order in $T_1T_2$ as
\eqn\leadorder{ Z_{\cal D}(t) \sim {1\over \kappa_{hh}(-t)^{T_1T_2}} \sim
\kappa_0^{-T_1T_2} \times \big(1+{\kappa_1\over \kappa_0} (t_c-t)^{5\over
6}\big)^{-T_1T_2}  }
Letting the time lapses $T_i$ scale like $T_i={\tau_i\over a}$, $a$ a 
small parameter, we see that we must approach the critical point as
\eqn\approa{ t=t_c(1-a^{12\over 5} \Lambda) }
where $\Lambda$ is the renormalized fugacity per tetrahedron, conjugated to the
renormalized volume ${\cal V}= V a^{12/5}$.
We deduce that for large $T_i$ 
\eqn\compV{ V \sim (T_1T_2)^{6 \over 5} }
hence the fractal dimension of the $(1+2)$-D Lorentzian
tetrahedral complexes is $d_F=12/5$.
In other words, the transverse area $T_1T_2$ explored by the tetrahedral complexes 
with $V$ plaquettes scales like $T_1T_2 \sim V^{5\over 6}$, to be
compared with the vertical length $T$ of triangulations with $A$ 
lozenges scaling like
$T\sim A^{1\over 2}$.  

\newsec{Relation with directed-site lattice animals}

It was shown by Dhar in ref.\DHAR\ that $(d+1)$-dimensional
problems of directed-site animals enumeration (DSAE) were related to 
$d$-dimensional nearest neighbor exclusion models.
The latter include the hard-dimer model (in $d=1$ dimension)
and the hard-hexagon model (in $d=2$ dimensions), respectively related
to DSAE on the simple square ($d+1=2$) and simple cubic ($d+1=3$) lattices.
More precisely, the generating functions of DSAE were expressed
as the density of occupation of the corresponding exclusion models, 
which in turn reduce precisely 
in the hard dimer and hard hexagon cases 
to the ratios yielding $Z^{C_0}_T(t)$ and $Z^{C_0}_{\cal D}(t)$ 
(see eqns. \corr\ and \formicorr). 
This strongly suggests the existence of a direct connection between 
semi-random Lorentzian lattices and directed-site lattice animals. 
In this section we will indeed establish an equivalence between 
the two problems.
As an outcome, this will give an alternative
derivation of the results of \DHAR\ in a more direct way. 
In the following,
we will focus for simplicity
on the cases of $(1+1)$-D Lorentzian triangulations and 
$(1+2)$-D Lorentzian tetrahedral complexes.   

\subsec{From Lorentzian triangulations to square lattice 
directed animals}

\fig{The equivalence between Lorentzian triangulations
with a fixed left projection (with edges on even time slices)
and directed site animals on the square lattice. We first build from the
configuration its skeleton obtained by squeezing all blocks into 
single edges (a). We then place the remaining edges of the skeleton
onto a square lattice tilted by $45^\circ$ by pushing each edge
to the leftmost available position on the lattice (b). This is equivalent to 
placing the successive left projections of the skeleton onto 
regularly spaced vertical lines here labeled $1,2,3,...$. 
We have also indicated
by arrows the two successors (in the next plane) of a vertex of the 
square lattice. A site on line $i$ can be occupied by an edge only if one
of its two predecessors (on line $i-1$) is occupied. 
The occupied sites form a directed site animal on the tilted square lattice
(c).}{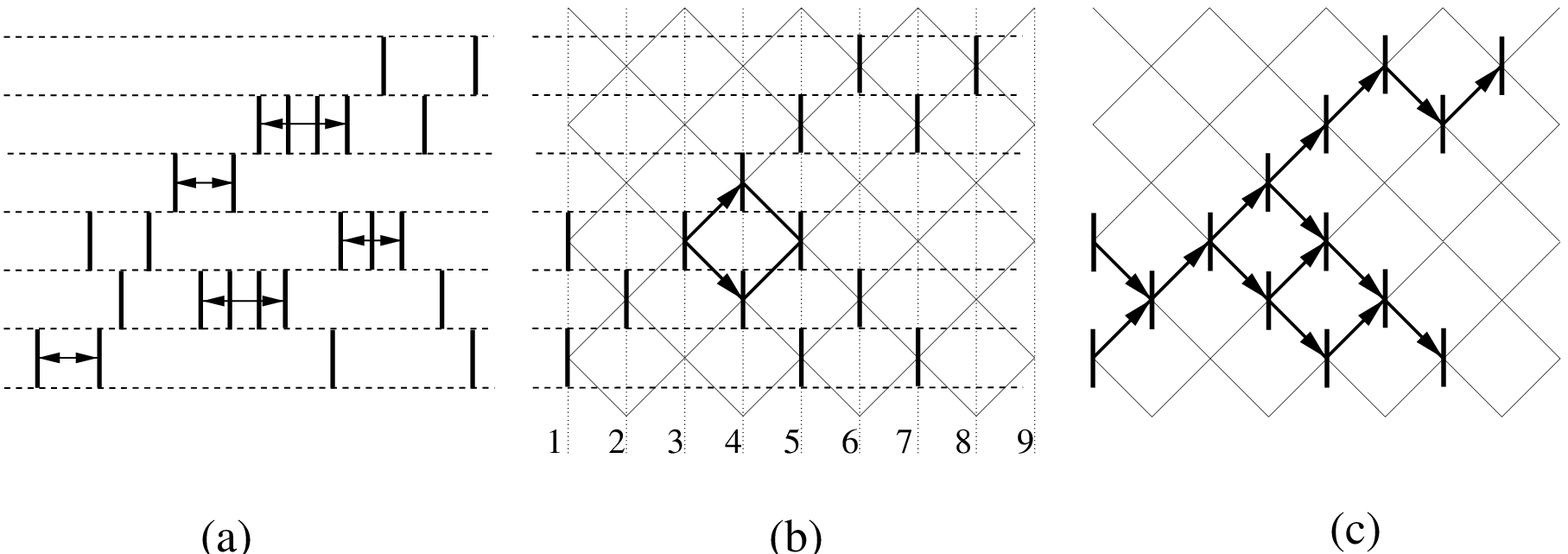}{13.cm}
\figlabel\skeleton

Let us start with the set of configurations of vertical edges of 
an arbitrary Lorentzian triangulation, with a fixed left projection $C$
made of hard dimers all occupying even time-slices. For each such
configuration we first construct its skeleton, obtained as follows.
We first decompose the edge configuration into blocks made of consecutive
edges within the same time slice that are not separated by
edges from neighboring slices (see Fig.\skeleton\ (a)).    
Squeezing each block into a single edge, we arrive at the skeleton of
the configuration, itself a particular edge configuration with projection
$C$, with no two consecutive edges within the same slice, as illustrated in
Fig.\skeleton\ (b).
Summing over all weighted configurations sharing the same skeleton
simply amounts to assign an effective fugacity $x=t/(1-t)$ for
each edge of the skeleton. Indeed, $x$ is nothing but the sum over
all block sizes $n=1,2,...$ with a weight $t^n$. 
We may now arrange the edges of the skeleton 
by representing its successive left projections along
regularly spaced vertical lines numbered $1$, $2$, $3$, ... 
(distant by $\sqrt{2}$, see Fig.\skeleton\ (b)).
Noting that successive projections alternate
between odd and even positions for all the edges, we see that they now
lie on a regular square lattice (tilted by $45^\circ$).
Moreover, the position of the edges satisfy the directed
lattice animal constraint that a site on the vertical line $i$
can be occupied only if one of its immediate neighbors on the lattice
at vertical line $i-1$ is occupied (see Fig.\skeleton\ (c)). 
In conclusion, the skeletons
of configurations with projection $C$ are in one-to-one
correspondence with directed lattice animals on the square
lattice with the same set $C$ as ``source".  
We therefore end up with an identity between the generating function 
$A_C(x)$ of square lattice animals with source $C$ and activity
$x=t/(1-t)$ per occupied site and the partition function of
Lorentzian triangulations with activity $t$ per dual edge and fixed left
projection $C$. 
\eqn\latan{ A_C(x)= t^{|C|} \, Z^{(C)}(t), \ \ {\rm with}\ \ 
t={x \over 1+x}, }
where we have used the thermodynamic partition function
$Z^{(C)}(t)=\lim_{T\to \infty}Z_T^{(C)}(t)$, well-defined 
for any finite $C$.

The inversion relation \corr\ allows then to write $A_C(x)$ as the
density of occupation $\rho_C(z)$ of the set $C$ in the hard dimer model with
fugacity $z=-t=-x/(1+x)$, namely
\eqn\othanim{ A_C(x)= (-1)^{|C|} (-t)^{|C|} {Z^{hd}_{\IZ \setminus C}(-t)\over
Z^{hd}_{\IZ}(-t)} = (-1)^{|C|} \rho_C(-t) }
with $t=x/(1+x)$ as above. This coincides with the result of Dhar \DHAR.
Our derivation using left projections is slightly more direct than that of 
\DHAR, which uses an intermediate connection with a dynamical
crystal-growth model and relies on the {\it similarity} of master equations
rather than on a direct correspondence between configurations. Note
also that our inversion relation \funda\ 
is not a simple rephrasing of the connection between directed animals and
hard dimers, as it gives for instance access to the (finite size)
partition function of the model of Lorentzian triangulations, 
which is not a density of occupation and has no direct animal interpretation. 
Moreover, in this section,
we have artificially restricted our model to left
projections with only edges in even positions, which was needed for the animal
interpretation. However we have a Lorentzian surface interpretation for all
other choices of the left projection $C$, and a relation to hard
dimer densities too.

\subsec{From Lorentzian tetrahedral complexes to cubic lattice 
directed animals}

\fig{The possible positions (a) for plaquettes in successive
partial projections of the skeletons. Plaquettes in planes
$1,2,3,4$... lie in the sublattices $1,2,3,1$... The positions
form a simple cubic lattice whose cell (b) is represented together
with the label of the corresponding $[111]$ crystalline planes. 
The arrows in both pictures indicate the
six immediate successors (in the planes $2$ and $3$)
of a vertex of the plane $1$.
For a given skeleton, only part of the available positions are
filled by plaquettes. 
A position of the cubic lattice can be occupied by a 
plaquette of a skeleton only if one of its 
six predecessors is occupied too.}{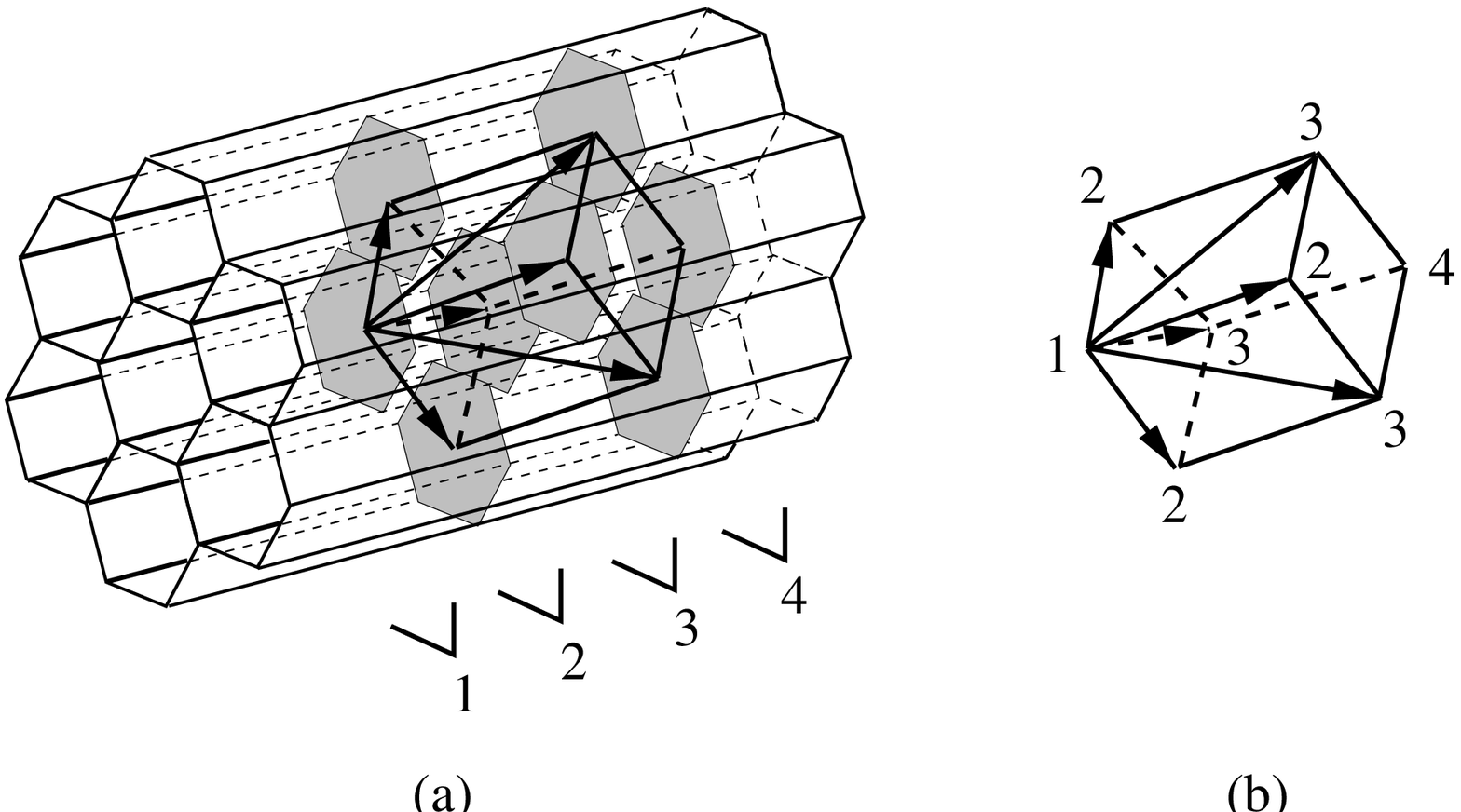}{13.cm}
\figlabel\animal

The above construction generalizes nicely to the case of Lorentzian
tetrahedral complexes, which are in one-to-one correspondence with
directed site lattice animals on the simple cubic lattice.
We may repeat the above construction. We start again from a plaquette
configuration with fixed projection $C$ made only of plaquettes
lying on one of the three triangular sublattices 
\foot{The triangular
lattice is naturally decomposed into three triangular sublattices,
such that no two sites of the same sublattice are adjacent.}  
numbered $1,2,3$
of the triangular lattice formed by the centers of the tubes.
The skeleton of the plaquette configuration is obtained again
by first decomposing it into blocks of successive plaquettes 
within the same tube, and shrinking each of these blocks into 
a single plaquette. Again, the partition function of the original
configurations with fixed projection $C$ and that of the skeletons
with projection $C$ are identified by assigning an effective fugacity 
$x=t/(1-t)$ to the plaquettes of the skeletons. 
The plaquettes of the skeleton are then placed at the vertices of
a simple cubic lattice by taking successive partial
projections as follows. 
We start from the left projection $C$ whose plaquettes lie on the
sublattice, say $1$, and place them on a first section plane
numbered $1$. The left projection of the rest of the skeleton
has plaquettes now lying on the sublattices $2,3$ only. We pick the
partial projection made of the plaquettes of this projection belonging
to the sublattice $2$, and place them on a second plane, numbered $2$,
parallel to number $1$, and distant by $1/\sqrt{2}$. Note that this
partial projection might be empty. We then go on by considering the
rest of the skeleton, whose left projection is now made 
of plaquettes on sublattices $3$ or $1$, out of which we retain those on the
sublattice $3$, etc ... Note that no two consecutive partial projection planes
can be empty.  
The parallel partial projection planes $1,2,3...$ are nothing but 
the successive $[111]$ crystalline planes of the simple cubic lattice
(see Fig.\animal\ (a)-(b)).
The skeleton may now be interpreted as a configuration 
where the vertices of the simple cubic
lattice may be occupied (or not) by plaquettes, with the directed site animal
constraint that a vertex in plane $j$ may be occupied only if one
of its preceding first or second nearest neighbors (respectively in 
planes $j-1$ and $j-2$) is occupied. 
This allows to identify the generating function for DSAE on the simple cubic lattice
with a fixed source $C$ and 
an activity $x$ per occupied site with the partition function 
of $(1+2)$-D Lorentzian  tetrahedral complexes with left projection $C$
and a weight $t=x/(1+x)$ per 
diamond-shaped building block (i.e. $t^{1\over 6}$ per tetrahedron).

This in turn allows to recover the relation between DSAE on the simple cubic lattice
and the hard hexagon model in two dimensions \DHAR, in a slightly more direct manner.

\newsec{Transfer matrices and integrability}

In ref.\FGKBIS\ it was shown that the existence of a transfer matrix formulation
of pure Lorentzian triangulations was instrumental for
deriving an effective one-dimensional (Calogero) Hamiltonian for the corresponding
continuum scaling limit.
It turns out that all the models introduced so far also have a more involved
but similar transfer matrix formulation in terms of the original Lorentzian 
surfaces.
In this section, we first show how to construct these transfer matrices,
using in particular our inversion formula. We then show how these matrices 
can be regarded as particular points of integrable families of commuting
matrices.

\subsec{Transfer matrices for Lorentzian surfaces}

\fig{The transfer matrix of Lorentzian triangulations (a) connecting
$i$ lower half-edges to $j$ upper ones across a given time-line.
The corresponding (non-empty) hard-dimer configurations (b) and their
respective weights.}{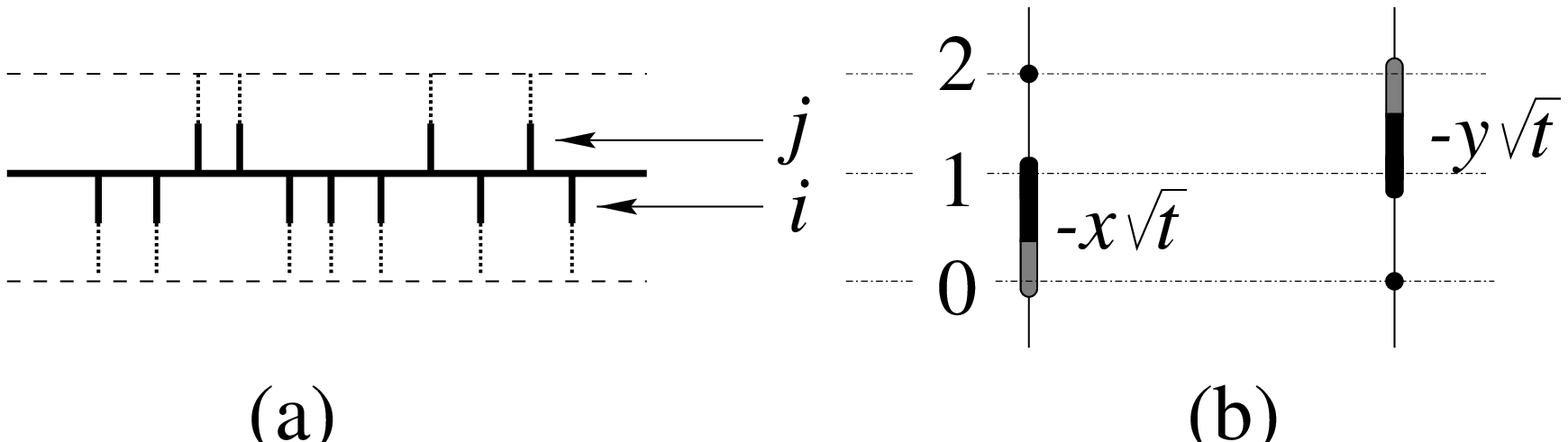}{11.cm}
\figlabel\transpur

Let us first recall the form of the (infinite)
transfer matrix for Lorentzian triangulations.
This matrix $T(t)$ transfers a row of half-edges across a time-line into another
row of half-edges (see Fig.\transpur\ (a)). A state is simply characterized by the
number $i$ (resp. $j$) of lower (resp. upper) half-edges, and the matrix element
simply counts the number ${i+j\choose i}$ 
of ways of arranging these half-edges along
the time-line, together with a weight $\sqrt{t}$ per half-edge. 
Another way of obtaining this matrix $T(t)$ consists in computing
its generating function $\Theta_1(x,y\vert t)=\sum_{i,j\geq 0} x^i y^j T(t)_{i,j}$
by use of our inversion formula \funda, namely by expressing it as the inverse
of the partition function of hard dimers on $[0,2]$, with weights $-\sqrt{t}x$
(resp. $-\sqrt{t} y$) per dimer in the slice\foot{We denote
by $m$ the slice $[m,m+1]$.} $0$ (resp. $1$) (see Fig.\transpur\ (b)):
\eqn\genpur{ \Theta_1(x,y\vert t)={1 \over 
\pmatrix{1 & {\rm i}y} \pmatrix{ 1 & {\rm i} \sqrt{t} \cr
{\rm i} \sqrt{t} & 0\cr} \pmatrix{ 1 \cr {\rm i}x \cr} }={1\over 1-\sqrt{t}(x+y)}
=\sum_{i,j\geq 0} {i+j\choose i} t^{i+j\over 2} x^i y^j }
Note that we have used a symmetrized version of the transfer matrix for
hard dimers on the line, differing from that of \tmathd\ by a simple conjugation.
The factors of i ensure that the dimers get the correct negative weight.
This is appropriate to account for the fact that we count only half of the edge
weight for the half-edges. With this definition, we have a similar formula
for the $T$-th power of $T(t)$ generated by $\Theta_T(x,y\vert t)$, namely 
\eqn\genpurT{ \Theta_T(x,y\vert t)={1 \over
\pmatrix{1 & {\rm i} y} \pmatrix{ 1 & {\rm i} \sqrt{t} \cr
{\rm i} \sqrt{t} & 0\cr}^T \pmatrix{ 1 \cr {\rm i} x \cr} }}
corresponding to the inverse of the partition function of hard dimers
on $[0,T+1]$ with a weight $-t$ per dimer on the slices $1,2,..,T-1$
and weights $-\sqrt{t} x$ (resp. $-\sqrt{t} y$) in the slice $0$ (resp. $T$).
This quantity is simply related to the loop-loop propagator \loplop\ by
$\Theta_T(x,y\vert t)=G_{T+1}(x/\sqrt{t},y/\sqrt{t}\vert t)$.

\fig{The transfer matrix of Lorentzian surfaces made of time-like
hexagons i.e. with dual edges of fixed length $k=2$ (a) connecting
lower and upper half-edges across a given time-line. The lower 
and upper half-edges
each come in two types according to their position ($1$ or $2$)
along the full edge.
The corresponding (non-empty) hard-trimer configurations (b) and their
respective weights.}{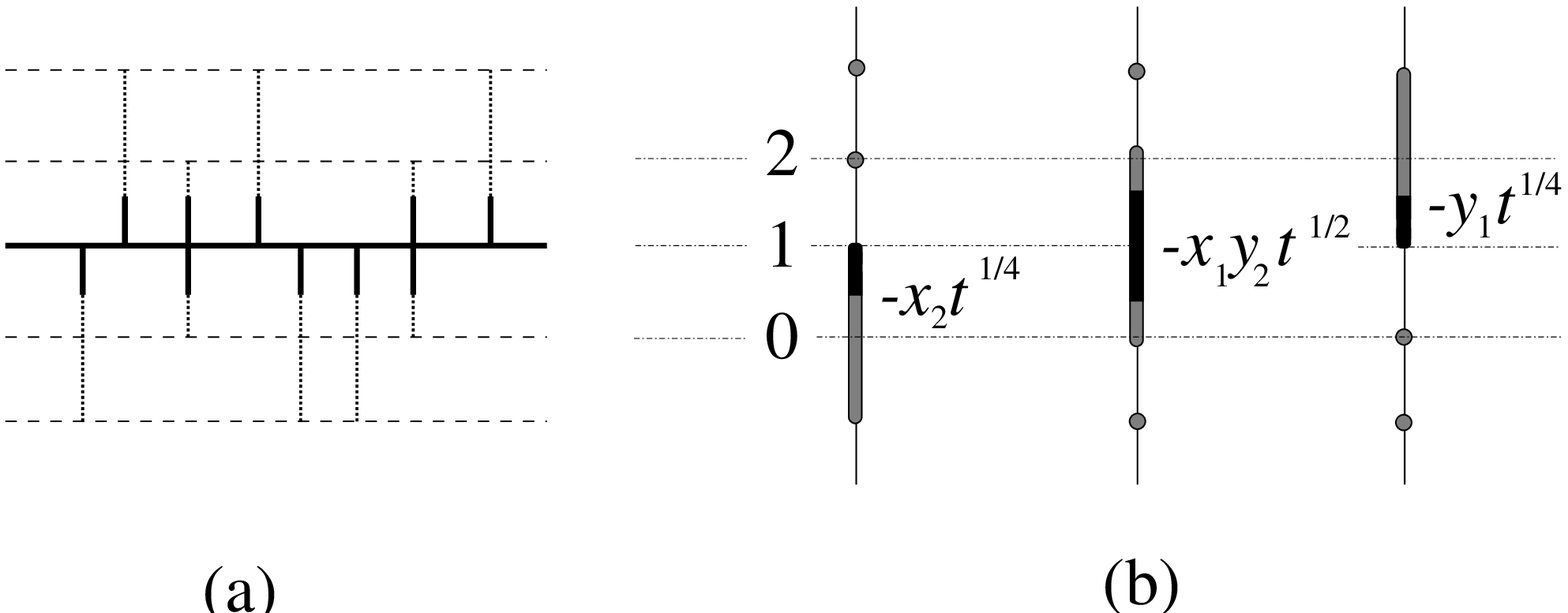}{13.cm}
\figlabel\transfer

Let us now turn to the model of Lorentzian surfaces
made of $2(k+1)$-gons defined in  Section 2.3.
To get a nice row-to-row transfer matrix $T_k(t)$, we must distinguish
between the possible states in which a given half-edge can be
according to its relative positioning along the full edge of size $k$ it
belongs to (see Fig.\transfer\ (a)). 
Let $i_r$ (resp. $j_r$) denote the total numbers of 
lower (resp. upper) half-edges occupying the position $r=1,2,...,k$ along
the edges of length $k$ they belong to. The corresponding transfer 
matrix elements read
\eqn\tmatfuscat{ T_k(t)_{i_1,...,i_k;j_1,...,j_k}= t^{{1\over 2k}
\sum_{r=1}^k (i_r+j_r)}
{j_1+\sum_{r=1}^k i_r \choose j_1} \prod_{r=1}^{k-1} \delta_{j_{r+1},i_r} }
expressing that a half-edge in position $r$ transfers into 
one in position $r+1$
for $r=1,...,k-1$ whereas one must choose the position of the $j_1$ {\it new}
upper half-edges wrt all the other previously existing ones, with a 
weight $t^{1\over 2k}$ per half-edge.
Note that for $k>1$ this transfer matrix is no longer symmetric.
As before, we use the inversion formula \fundafin\ to write the free
boundary partition function
$\Theta_T^{(k)}(x_1,...,x_k;y_1,...,y_k\vert t)$ over a time
lapse  $T$ as the inverse of the partition function of hard $(k+1)$-mers on
$[0,T+1]$ possibly cut on the time extremities, with a weight $-t$ per 
$(k+1)$-mer visiting the slices $1,2,..,T-1$,
and a weight $-x_r t^{2k-2r+1\over 2k}$ (resp. $-y_r t^{2r-1\over 2k}$)
for a $(k+1)$-mer cut in position $r$ in the slice $0$ (resp. $T$), for $T\geq k$.
When $T<k$, we must pay attention to $(k+1)$-mers that are cut at both extremities:
these receive a weight $-x_r y_{T+r} t^{T\over k}$, $r=1,2,...,k-T$ (see
Fig.\transfer\ (b)).  
This finally leads to
\eqn\invfusT{\eqalign{
\Theta_T^{(k)}(x_1,...,x_k &;y_1,...,y_k\vert t)= 
{1 \over w_k^t(y) \big({\cal T}_k(t)\big)^T v_k(x) } \cr
{\rm with} \ \ w_k^t(y)&=\pmatrix{1 & \alpha^{2k-1} y_1 
& \alpha^{2k-3} y_2 & \cdots & 
\alpha^3 y_{k-1} & \alpha y_k } \cr
v_k^t(x)&=\pmatrix{1 & \alpha x_1 & \alpha^3 x_2 
& \cdots & \alpha^{2k-3} x_{k-1} & 
\alpha^{2k-1} x_k } \cr
{\cal T}_k(t)&=\pmatrix{ 1 & 0 & 0 & \cdots & 0 & \alpha t^{1\over 2k}\cr
\alpha t^{1\over 2k} & 0 & 0 & \cdots & 0 & 0\cr
0 & \alpha^2 t^{1\over k} & 0 & \cdots & 0 & 0 \cr
0 & 0 & \alpha^2 t^{1\over k} & \cdots & 0 & 0 \cr
\vdots & \vdots & \vdots & \ddots & \vdots & \vdots \cr
0 & 0 & 0 & \cdots &\alpha^2 t^{1\over k} & 0 \cr} \cr}}
where we have set $\alpha=e^{{\rm i} \pi \over 2k}$.
The states on which ${\cal T}_k(t)$ acts can be understood as being
respectively
the vacuum, the first (lowest) elementary segment of the $(k+1)$-mer,
the second, third,..., $k$-th one. 
As before, the phases in \invfusT\
ensure that all multimers (including the cut ones) have
a minus sign in front of the fugacity. 
Note also the explicit ``up-down" symmetry 
of ${\cal T}_k(t)$ implemented by the symmetric matrix
${\cal R}_{i,j}=\delta_{j,k+1-i}$ for $i,j\geq 1$ and 
${\cal R}_{0,j}=\delta_{j,0}$, 
namely that  the matrix ${\cal R} {\cal T}_k(t)$ is symmetric
and that ${\cal R}v_k(x) = w_k(x)$.
This translates into the symmetry of $\Theta_T^{(k)}$ under the interchange of
$x_r \leftrightarrow y_{k+1-r}$, $r=1,2,...,k$.
A preliminary remark is in order: if we take the formula \invfusT\ for $T=0$,
we get
\eqn\tzero{ \Theta_0^{(k)}(x_1,...,x_k ;y_1,...,y_k\vert t)=
{1\over 1-\sum_{r=1}^k x_r y_r}=\sum_{i_1,...,i_k\geq 0}
{(\Sigma i_r)! \over \Pi i_r!} \prod_{r=1}^k (x_ry_r)^{i_r}}
which generates the diagonal matrix $P$ with diagonal
entries $P_{i_1,...,i_k}={(\Sigma i_r)! \over \Pi i_r!}$ rather than just 
the identity. More generally, $\Theta_T^{(k)}$  is the generating function for
$\big(T_k(t)\big)^T P$, where the boundary operator  
$P$ corresponds to a preliminary ordering
of the half-edges of the initial state over which $T_k(t)$ acts. Note that
the action of $T_k(t)$ automatically orders the final state. 
In particular, for $T=1$  
the result \invfusT\ reduces to
\eqn\propfusc{\eqalign{&\Theta_1^{(k)}(x_1,...,x_k;y_1,...,y_k\vert t)=
{1\over 1-t^{1\over 2k}(x_k+y_1)-t^{1\over k}(x_1y_2+x_2y_3+...+x_{k-1}y_k)} \cr
&=\sum_{i_1,...,i_k,j_1\geq 0} t^{j_1+i_k+2\Sigma_{1\leq r\leq k-1} i_r\over 2k}
(x_1y_2)^{i_1}...(x_{k-1}y_k)^{i_{k-1}} x_k^{i_k} y_1^{j_1}
{(j_1+\Sigma_{1\leq r\leq k} i_r)!\over j_1! i_1!i_2! ...i_k!}\cr}}
which is nothing but the generating function of $T_k(t) P$, with $T_k(t)$
as in \tmatfuscat.  
The up-down symmetry
is recovered by noting that $R T_k(t) R = P^{-1} T_k(t)^t P$, 
where $R$ is the matrix
with entries $R_{i_1,...,i_k;j_1,...,j_k}=\prod_{r=1}^k \delta_{j_r,i_{k+1-r}}$
corresponding to the interchange $x_r\leftrightarrow x_{k+1-r}$, while the
transposition expresses the interchange of $x_r$ and $y_r$.

More generally, the above construction can be generalized to the
hard multimer models of Section 2.2, leading to an explicit transfer matrix
formulation of the corresponding generalized Lorentzian surfaces. 
Details are gathered in Appendix A below. 
Finally, let us mention that the above extends straightforwardly to higher
dimensions. 
In the case of $(1+2)$-D Lorentzian tetrahedral complexes for
instance,
we may define a ``plane-to-plane" transfer matrix acting in the
second time direction ($T_2$, see Fig.\briques) and transferring from a
$(1+1)$-D Lorentzian triangulation to the one above (see Fig.\intersec). 
The generating function
for this transfer matrix (multiplied by a diagonal 
ordering operator $P$ as
before) is expressed as the inverse of a decorated partition function
for hard hexagons via our inversion formula. The latter is itself the
generating function for the matrix elements of the row-to-row transfer 
matrix for hard hexagons \BAXHH.

\subsec{Integrability and the inversion relation}

In ref.\FGK\ it was observed that the model of pure Lorentzian triangulations
was a point ($a=1$) along an integrable family of models including both triangles
and squares (with two time-like and two space-like edges), 
with weights $a \sqrt{t}$ per triangle and $u=t(1-a^2)$ per square. 
The corresponding surfaces are generated in the dual picture
by means of a row-to-row transfer matrix $T(t,a)$ whose element
$T(t,a)_{i,j}$ corresponds to the transfer from $i$ lower vertical
edges to $j$ upper ones across a time-line, $i,j\geq 0$. 
It was shown that the transfer matrices $T(t,a)$ and $T(t',a')$ 
commute provided 
\eqn\comm{ {1-t (1-a^2) \over a \sqrt{t}} = {1-t' (1-a'^2) \over a' \sqrt{t'}} }
In our language this model with triangles and squares corresponds to
having $2(k+1)$-gons of arbitrary size $k=1,2,3,...$ made of the piling up of
$(k-1)$ squares across $k$ consecutive time-slices, and terminated by two triangles
on the top and bottom of the pile. These objects come with a weight
$t_k=a^2 t u^{k-1}$, $k=1,2,3,...$.
We immediately deduce from our general inversion formula that the partition function
$\mu(t,a)$ for these surfaces with left staircase boundary conditions and with
semi-infinite time range $[0,\infty)$ satisfies the equation \charmut\
\eqn\equamu{ \mu= 1+ \sum_{k=1}^\infty a^2 t u^{k-1} \mu^{k+1} \ \
\Rightarrow \ \ t\mu^2=(1+u) \mu -1 }
Upon the rescaling ${\tilde \mu} = (1+u) \mu $ and ${\tilde t}= t/(1+u)^2$ we
end up with the same equation as that of the pure Lorentzian triangulation case $a=1$
($u=0$). 
As a non-trivial outcome of our formulation, the integrability condition for
the infinite size transfer matrices
$T(t,a)$ can be deduced from a similar condition on the $2\times 2$ transfer matrices
of the corresponding hard multimer models. Let us write the generating function
$\Theta_2(x,y\vert a,t,a',t')=\sum_{i,j\geq 0}x^i y^j (T(t',a')T(t,a))_{i,j}$
of the product $T(t',a')T(t,a)$ as the inverse of a multimer
partition function on $[0,2]$ with weights $-a\sqrt{t} x$ (resp. $-aa'\sqrt{tt'}$
and $-a'\sqrt{t'} y$) for dimers in slice $0$ (resp. $1$ and $2$), $-a'\sqrt{t'}ux$
(resp. $-a\sqrt{t} u' y$) for trimers covering the slices $0,1$ (resp. $1,2$) and finally
$-x y u u'$ for quadrimers covering all the slices. This reads
\eqn\redgtre{ \Theta_2(x,y\vert a,t,a',t')=
{1 \over \pmatrix{1 & {\rm i} y} \pmatrix{ 1 & {\rm i} a'
\sqrt{t'} \cr
{\rm i} a' \sqrt{t'} & u'\cr} \pmatrix{ 1 & {\rm i} a \sqrt{t} \cr 
{\rm i} a \sqrt{t} & u\cr}
\pmatrix{ 1 \cr {\rm i} x \cr} } }
Similarly the generating function $\Theta_2(x,y\vert a',t',a,t)$ is expressed by the
same formula with the two $2\times 2$ matrices exchanged. The commutation of the
original transfer matrices translates therefore into that of these two
$2\times 2$ ones. A simple calculation shows that they commute provided
$(1-u)/(a\sqrt{t})= (1-u')/(a'\sqrt{t'})$, which is precisely the integrability
condition \comm.

This suggests to look for integrable families of finite size matrices
corresponding to decorated multimer models and to interpret them as integrable
models of more involved Lorentzian surfaces. 
For simplicity, let us consider the case of symmetric 
$(k+1)\times (k+1)$ matrices.
For illustration, we may consider
a model of colored hard dimers with a color index $i=1,2,...,k$,
and a weight $z_i$ for each dimer of color $i$,
with transfer matrix 
\eqn\celem{{\cal T}_0
=\pmatrix{1 & \sqrt{z_1} & \sqrt{z_2} & \cdots & \sqrt{z_k}\cr
\sqrt{z_1} & 0 & 0 &\cdots & 0 \cr
\sqrt{z_2} & 0 & 0 & \cdots & 0 \cr
\vdots & \vdots & \vdots & & \vdots \cr
\sqrt{z_k} & 0 & 0 & \cdots & 0 \cr} }
This is nothing but the original hard dimer 
model with fugacity $z=z_1+z_2+...+z_k$.
However, we now want to view it as a particular point of an integrable 
family of transfer
matrices ${\cal T}$, depending on $k(k+3)/2$ new 
parameters ${\cal T}_{i,j}={\cal T}_{j,i}$,
$i,j=0,1,2,...,k$ where we fix ${\cal T}_{0,0}=1$. 
The parameters ${\cal T}_{0,j}=\sqrt{w_j}$, $j=1,2,...,k$ are interpreted as
new fugacities per dimer of color $j$.
The numbers ${\cal T}_{i,j}$ for $i,j\geq 1$ can be interpreted as
follows. We first allow the dimers to pile up so as to form
multicolored multimers of length $m$, and we assign a weight 
$\sqrt{w_{i_1}w_{i_m}} \prod_{p=1}^{m-1}
{\cal T}_{i_p,i_{p+1}}$ for each $(m+1)$-mer formed by a sequence of dimers
of colors $i_1,i_2,...,i_m$. 
To get an integrable family, we must impose that all the ${\cal T}$
share the same orthonormal basis of eigenvectors as ${\cal T}_0$ (this
involves fixing a $(k-1)$-dimensional rotation acting on the kernel 
of ${\cal T}_0$).
Denoting by $\psi_i^{(m)}$ the $i$-th entry of the $m$-th eigenvector
of ${\cal T}_0$, and by $\Lambda_m$ the corresponding
$m$-th eigenvalue of ${\cal T}$, for $i,m=0,1,2,...,k$,  we have 
\eqn\wriT{ {\cal T}_{i,j}= \sum_{m=0}^k \Lambda_m \psi^{(m)}_i 
\psi^{(m)}_j } 
where the $\Lambda_m$'s are free parameters satisfying the constraint
\eqn\constraint{{\cal T}_{0,0}=1=\sum_m \Lambda_m ( \psi^{(m)}_0)^2}
This shows that the matrices $\cal T$ form generically a
$k$-parameter family passing by the point ${\cal T}_0$ (with $w_i=z_i$).
This also gives a parametrization of the dimer factors 
${\cal T}_{0,i} = \sqrt{w_i}=\sum_m \Lambda_m \psi^{(m)}_0
\psi^{(m)}_i$ in terms of the $k+1$ eigenvalues $\Lambda_m$. 
Let us display the case $k=1$ for illustration.
We first  
diagonalize the matrix ${\cal T}_0=\pmatrix{1 & {\rm i} \sqrt{t} \cr
{\rm i}\sqrt{t} & 0\cr}$ with the resulting eigenvectors:
\eqn\eigenpsi{ \psi^{(0)\, t}= {1\over \sqrt{1-q^2}} 
\big(1 , {\rm i} q \big),
\qquad \psi^{(1)\, t}= {1\over \sqrt{1-q^2}} \big({\rm i} q , -1\big)}
where we have set $1/\sqrt{t}=q+1/q$, and $q<1$.
The one-parameter integrable family takes the form
\eqn\integTr{ {\cal T}_{i,j}= (1+{\rm i} q\sqrt{w}) \psi_i^{(0)}\psi_j^{(0)}
+(1+{\rm i}\sqrt{w}/q)  \psi_i^{(1)}\psi_j^{(1)} }
where we have parametrized the eigenvalues $\Lambda_0$ and $\Lambda_1$,
satisfying $\Lambda_0 ( \psi^{(0)}_0)^2+ \Lambda_1 ( \psi^{(1)}_0)^2=1$
in terms of the matrix element ${\cal T}_{0,1}=\sqrt{w}$.
Eqn.\integTr\ coincides with ${\cal T}=\pmatrix{1 & {\rm i} a\sqrt{t}\cr
{\rm i} a\sqrt{t} & u \cr}$ provided $\sqrt{w}={\rm i} a \sqrt{t}$ 
and $u=t(1-a^2)$.

A slight generalization of the previous case allows to include the general
Lorentzian surfaces models of Sections 2.2 and 2.3 in larger integrable families.
As opposed to the previous case, the corresponding 
transfer matrix ${\cal T}_0$ is no longer symmetric 
(see for instance eqn.\invfusT).
Still, the ``up-down" symmetry of the problem 
allows to make the transfer matrix symmetric, by use of
a certain involution $i\to r(i)$ of its indices,
implementing the up-down reflection of the multimers, while preserving
the vacuum ($r(0)=0$). 
More precisely,  let $\cal R$ be the matrix with entries
${\cal R}_{i,j}=\delta_{j,r(i)}$, 
then the up-down symmetry of ${\cal T}_0$ translates into the
fact that ${\cal R} {\cal T}_0$ is symmetric.
We now may look for an up-down symmetric
integrable deformation ${\cal T}$ of ${\cal T}_0$,
namely such that ${\cal R}{\cal T}$
is symmetric, and moreover satisfying the condition that
${\cal T}_{0,0}=1$.
Denoting again by $\psi_i^{(m)}$ a basis of diagonalization of ${\cal T}_0$, 
we first note that by virtue of up-down symmetry 
it can be normalized so as to be ``${\cal R}$-orthonormal", 
namely that $\big({\cal R} \psi^{(m)}\big)^t \psi^{(p)}=\delta_{m,p}$.
We still write the family ${\cal T}$ in a form similar to \wriT, 
taking into account the up-down symmetry, namely
\eqn\newriT{ {\cal T}_{i,j}= \sum_{m} \Lambda_m \psi^{(m)}_i
\psi^{(m)}_{r(j)} }
where the $\Lambda$'s satisfy the same constraint as before:
${\cal T}_{0,0}=1=\sum_m \Lambda_m (\psi_0^{(m)})^2$. 
The case of hard $(k+1)$-mers of Section 2.3 is obtained by taking
${\cal T}_0\equiv {\cal T}_k(t)$ of eqn. \invfusT, with the same
up-down symmetry matrix ${\cal R}_{i,j}=\delta_{j,k+1-i}$ for
$i,j=1,2,...,k$ and ${\cal R}_{0,j}=\delta_{j,0}$.  
The case of hard multimers
of Appendix A corresponds to taking 
${\cal T}_0\equiv {\cal T}^{(hm)}$, and
the up-down symmetry matrix ${\cal R}={\cal R}^{(hm)}$. 

In conclusion, we have shown that all the $1$D hard-multimer
transfer matrices used in
this paper are part of larger integrable families corresponding to
the introduction of extra interactions between multimers. 
This in turn implies that all the transfer matrices for 
Lorentzian surfaces defined in  Section 2
are also part of larger integrable families allowing for the original
polygonal tiles to form larger composite objects, with new weights. 
This extends nicely to the case of $(1+2)$-D Lorentzian 
tetrahedral complexes as well. Indeed, as noted by Baxter \BAXHH, the
hard hexagon model is part of a larger integrable family
(hard square model with diagonal interactions). Once translated
in the language of plaquettes this amounts to allow for hexagons
to coexist on neighboring sites along one of the three directions
of the triangular lattice of tubes, 
with a new contact Boltzmann weight.
The corresponding transfer matrices form via our inversion relation
the desired integrable family of models, describing complexes
in which the diamond-shaped building blocks 
can be glued together along that same direction so as to form 
larger straight objects of arbitrary length, with
new Boltzmann weights.

\newsec{Discussion and conclusion}

In this paper, we have first investigated general models of $(1+1)$-D 
Lorentzian surfaces. Viewing the pure Lorentzian triangulations  
as semi-random surfaces made only of time-like lozenges, we have
considered semi-random surfaces made of larger time-like tiles, namely
obtained by piling up lozenges, thus forming $2(i+1)$-gons, $i=1,2,3...$.
All these models have been solved by use of an inversion relation 
expressing their partition functions in terms of
those of associated hard multimers in $1$D.
We have then showed how to obtain multicritical models of these
generalized Lorentzian surfaces by an appropriate fine-tuning of the activities
of the various tiles.
These models involve only time-like interactions, a key point for the 
inversion relation to $1$D hard objects to hold. Indeed, this relation relies
entirely on the fact that the only horizontal (space-like) coupling is
through the no-crossing prescription of edges.
It would be interesting however to try to build more involved inversion
relations, possibly using horizontal (space direction) transfer matrices
between successive vertical projections, and use them to solve models
with horizontal (space-like) as well as vertical (time-like) interactions, 
such as the Ising model on Lorentzian surfaces.

We then generalized our construction so as to build a new 
higher-dimensional model of $(1+2)$-D Lorentzian tetrahedral complexes, 
with two regular
(time-like) and one random (space-like) directions. This model was solved by
using a generalized higher-dimensional version of the previous inversion formula,
relating it to the $2$D hard hexagon model solved by Baxter \BAXHH.
This model is the most natural extension of that of Lorentzian triangulations
to higher dimension with tetrahedra playing the role of triangles. We also showed
how to interpret the model as the time evolution of Lorentzian triangulations
in successive layers. In this respect, we obtain a toy-model for $(2+1)$-D Lorentzian
gravity, where the space-like triangulations are restricted to be themselves 
Lorentzian.
It is interesting to note that other plaquette models can be similarly constructed
by considering tubes with sections forming different lattices. For instance, we may 
consider a model of square plaquettes with section the square lattice or 
of triangular plaquettes with section the triangular lattice. In an analogous
way, the latter are respectively related to the hard square and hard triangle 
models in two dimensions. Unfortunately, these have not yet been solved exactly,
although very precise results are known about them \BAX.
More importantly, when we try to interpret these as models of simplicial
Lorentzian manifolds, we find that the corresponding elementary simplices (replacing
the tetrahedra) become degenerate objects with pairs of vertices linked
by more than one edge, that are usually discarded in simplicial gravity.
This is why we did not go into details of these models.
Another possible generalization is to keep hexagonal tubes and tetrahedra as elementary
objects, but to form larger building blocks than the diamond-shaped
dodecahedra considered so far. These larger objects are obtained by gluing several
dodecahedra in the two time directions in the same way as we did for lozenges.
Models based on these building blocks with say activities $t_i$ for the $i$-th
type of block are still  related to models of hard objects on the triangular lattice,
namely with plaquettes obtained by gluing hexagonal plaquettes together, and 
attaching to them activities $z_i=-t_i$. 
These $2$D models seem not to have been considered yet (in particular, they are 
not identical to the so-called
ABF models \ABF\ generalizing the hard hexagon model). 
Still it is very plausible that by again fine-tuning the activities $t_i$ one
should be able to reach multicritical points. It is natural to believe
that these correspond to non-unitary conformal field theories with central
charges $c(2,2m-1)=1-3(2m-3)^2/(2m-1)$, where the case of hard hexagons
corresponds to the Lee-Yang edge singularity with $m=3$, $c=-22/5$. 
The corresponding thermal dimension is known to be
$h_{1,3}=(5-2m)/(2m-1)$ henceforth the singularity of the free energy 
has the critical exponent $2-\alpha=1/(1-h_{1,3})= (2m-1)/(2(2m-3))$.
The corresponding fractal dimension of the associated multicritical 
Lorentzian manifolds would read $d_F=4(2m-3)/(2m-1)$.
 
Let us now point out that there is
an obvious natural higher-dimensional generalization
of tetrahedral complexes, now built with $d+1$-simplices living
in $d+1$-dimensional ``tubes" whose $d$-dimensional
section is the regular $d$-dimensional generalization of
the FCC lattice \TROISD, in correspondence with
a model of nearest neighbor exclusion on the vertices of that lattice.

Another direction of generalization consists in staying in low
dimension e.g. in $(1+1)$-D, but to introduce disorder in the form of
randomly distributed activities, but constant within each tube. 
Indeed, the inversion relation of this paper holds as well for each 
realization of the disorder, with fixed tube-dependent activities.
In the case of quenched disorder, where we must compute
the average $\overline{{\rm Log}\, Z}$
over all possible realizations of the disorder, we can use the
inversion formula term by term in the average, to write the
result as minus the quenched disorder average of the corresponding hard-object
model, namely $-\overline{{\rm Log}\, Z_h}$. Clearly, the inversion
relation does not allow to relate the two annealed disorder partition functions
$\overline{Z}$ and $\overline{Z_h}$ of the two problems. 

Another type of disorder consists in considering plaquette
models with fixed activities but defined on  
tubes whose section is itself arranged into random graphs, in connection to the
corresponding nearest neighbor exclusion problem on the same graphs. 
Again, the two quenched disorder problems are identical up to a sign.    

We may also consider tubes with section arranged into an arbitrary but
fixed graph. An interesting choice for such a graph is one where the
nearest neighbor exclusion model has already been solved, in
which case we can immediately convert the solution into that of the
corresponding Lorentzian plaquette model.  
One such example is the infinite rooted $q$-valent graph known as the Bethe
lattice, $q=2,3,...$. The corresponding plaquettes are $q$-gons, each connected
to one $q$-gon in the previous shell and to $q-1$ ones in the next. 
A simple calculation (see e.g. \BAX) shows that the thermodynamic free
energy of the nearest neighbor exclusion model $f_{nne}(z)$ per site,
with an activity $z$ per occupied site, reads 
\eqn\bethe{ 
f_{nne}(z)=(1-{q\over 2})\, {\rm Log}(2-\mu(-z)) -{q\over 2}\, {\rm Log}\, \mu(-z)}
where $\mu(t)$ is the Fuss-Catalan generating function \fussca\ with $k=q$,
obeying the equation $z=(1-\mu(t))/\mu(t)^q$, $t=-z$. For $q\geq 2$, the first
singularity of $f_{nne}(z)$ for $z<0$ occurs at the value $z=z_c$ such that 
$\mu(-z_c)=q/(q-1)$, resulting in a singularity with exponent $2-\alpha=1/2$,
translating into a fractal dimension $d_F=2$ for the corresponding Bethe
Lorentzian surfaces. 

Finally, the models considered here were shown to contain lattice animals
as a subclass. 
Conversely, we may apply this equivalence backwards to investigate
more refined properties of our semi-random lattices.
In particular, directed lattice animals are known to have two
distinct characteristic lengths, a longitudinal and a transversal one,
each with its own scaling exponent $\nu_\parallel$ and $\nu_\perp$.
While $\nu_\perp$ refers to our time correlation length exponent
(with $d_F=1/\nu_\perp$), it would be interesting to interpret
the second exponent in terms of our semi-random lattices.
On the other hand,
the animals interpretation holds as well for models with
$2(i+1)$-gons $i=1,2,3...$. However the animals corresponding for instance
to $(1+1)$-D models with edges of fixed length $k\geq 2$ are quite
unconventional, not to speak about those corresponding
to the multicritical models.  

After completion of this work, we became aware of
ref.\VIE\ where an analogous inversion formula
as that for $(1+1)$-D pure Lorentzian triangulations
was derived in the context of commutative monoids,
and of more recent developments on lattice animals
and exclusion models \BP\ \BM.

\vskip 2.cm

\noindent{\bf Acknowledgments} We thank T. Garel for pointing 
out to us ref.\DHAR, and J.-M. Luck for useful discussions.
We also thank D. Dhar for pointing out ref.\VIE\ to us
and M. Bousquet-M\'elou for further references.

\appendix{A}{Transfer matrix for general multimers}
 
Let us consider the general model of Lorentzian surfaces made
of $2(i+1)$-gons $i=1,2,...,k$ of Section 2.2. 
It is easy to see that the corresponding transfer matrix 
$T^{(hm)}$ is expressed in
terms
of a hard multimer one (denoted by ${\cal T}^{(hm)}$) through
a formula similar to \invfusT. The matrix ${\cal T}^{(hm)}$ is obtained by
superposing those for fixed $i$ (with respective transfer matrices    
$T_i(t)$ as in \invfusT) in the following manner. 
It must act on the following states:
the vacuum (label $0$), the unique dimer (label $1$), the first segment of a
trimer (label $2$), the second segment of a trimer (label $3$),..., the 
$j$-th segment of a $(i+1)$-mer (label $I(i,j)=j+i(i-1)/2$ for 
$j=1,2,...,i$),...
the $k$-th segment of a $(k+1)$-mer (label $k(k+1)/2$).
The matrix ${\cal T}^{(hm)}$ has therefore
total size $1+k(k+1)/2$ and its elements read
\eqn\elemT{\eqalign{
{\cal T}^{(hm)}_{0,0}&= 1\cr
{\cal T}^{(hm)}_{0,I(i,j)}&= {\cal T}_i(t)_{0,j} \quad i,j \geq 1 \cr
{\cal T}^{(hm)}_{I(i,j),0}&= {\cal T}_i(t)_{j,0} \quad i,j \geq 1 \cr 
{\cal T}^{(hm)}_{I(i,j),I(i,m)}&={\cal T}_i(t)_{j,m} \quad i,j,m \geq 1 \cr
{\cal T}^{(hm)}_{I(i,j),I(p,m)}&= 0 \qquad {\rm for}\ \ i \ne p \cr}}
where $T_i(t)$ is as in \invfusT. Note that this matrix is also ``up-down"
symmetric in the sense that ${\cal R}^{(hm)} {\cal T}^{(hm)}$ is symmetric,
where 
${\cal R}^{(hm)}$ is the symmetric matrix
with entries
\eqn\rsym{\eqalign{
{\cal R}^{(hm)}_{0,0}&=1\cr
{\cal R}^{(hm)}_{0,I(i,j)}&={\cal R}^{(hm)}_{I(i,j),0}=0\cr
{\cal R}^{(hm)}_{I(i,j),I(p,m)}&=\delta_{p,i}\delta_{m,i+1-j}\cr}}
for $1\leq j\leq i\leq k$.
The matrix ${\cal T}^{(hm)}$ leads through a formula analogous to \invfusT\
to the generating function
$\Theta_T^{(hm)}(\{x_{I(i,j)}\};\{y_{I(p,m)}\}\vert t_1,...,t_k)$
of $\big(T^{(hm)}\big)^T P^{(hm)}$ where the diagonal matrix
$P^{(hm)}$ implements the ordering of the initial state over which
$T^{(hm)}$ acts. More precisely, we have
\eqn\forgene{ \eqalign{
\Theta_T^{(hm)}&(\{x_{I(i,j)}\};\{y_{I(p,m)}\}\vert t_1,...,t_k)= 
{1\over 
w^t \big({\cal T}^{(hm)}\big)^T v } \cr
{\rm with} \ \ w_{I(i,j)}&=(w_i)_j(y_{I(i,j)}),\qquad w_0=1\cr
v_{I(i,j)}&=(v_i)_j(x_{I(i,j)}),\qquad v_0=1\cr}}
which for $T=0$ generates the elements of the diagonal
matrix ${P}^{(hm)}$ with diagonal entries:
\eqn\ordering{ {P}^{(hm)}_{\{i_{I(p,m)}\}} = {(\sum i_{I(p,m)})! \over
\prod i_{I(p,m)}!} }
while the formula \forgene\ leads for $T=1$ to the transfer matrix elements
\eqn\transmatgene{\eqalign{
T^{(hm)}_{\{i_{I(p,m)}\};\{j_{I(p,m)}\}}&= 
\prod_{p=1}^k
\left(
t^{{1\over 2p}\sum_{r=1}^p (i_{I(p,r)}+j_{I(p,r)})}
\prod_{r=1}^{p-1} 
\delta_{j_{I(p,r+1)},i_{I(p,r)}} \right)\cr
&\ \ \ \ \times
{\big(\sum_{p=1}^k (j_{I(p,1)} +\sum_{r=1}^p i_{I(p,r)})\big)! \over
\big(\sum_{1\leq r\leq p\leq k} i_{I(p,r)} \big)! 
\prod_{p=1}^k j_{I(p,1)}!} \cr}}
For illustration, the matrix ${\cal T}^{(hm)}$ reads for $k=2$:
\eqn\matonetwo{ 
{\cal T}^{(hm)}=\pmatrix{1 & {\rm i}\sqrt{t_1} & 0 & \alpha t_2^{1\over 4}\cr
{\rm i}\sqrt{t_1} & 0 & 0 & 0 \cr
\alpha t_2^{1\over 4}  & 0 & 0 & 0 \cr
0 & 0 & {\rm i}\sqrt{t_2} & 0\cr} }
where $\alpha=e^{{\rm i}\pi/4}$. This leads to the matrices
\eqn\matPT{\eqalign{
P^{(hm)}_{i_1,i_2,i_3}&= {(i_1+i_2+i_3)! \over i_1! i_2! i_3!}\cr
T^{(hm)}_{i_1,i_2,i_3;j_1,j_2,j_3}&= t_1^{i_1+j_1\over 2}
t_2^{2i_2+i_3+j_2\over 4} \delta_{j_3,i_2} 
{(i_1+i_2+i_3+j_1+j_2)! \over (i_1+i_2+i_3)! j_1! j_2!} \cr}}
where $i_1$ (resp. $i_2,i_3$) denote the total numbers of halves 
of single edges
(resp. halves of first, second segments of edges of length $2$) in the
lower part and similarly for the $j$'s in the upper part. Note that the
combinatorial factor in the second line of \matPT\ expresses the choice
of position of the two new types of edges (of respective length $1$ and $2$,
in numbers $j_1$ and $j_2$) wrt the already existing ones.   

\listrefs
\end